\documentclass[letterpaper,journal]{IEEEtran}

\usepackage{amsmath,amsfonts,amssymb}
\usepackage{algorithmic}
\usepackage{algorithm}
\usepackage{array}
\usepackage{textcomp}
\usepackage{url}
\usepackage{verbatim}
\usepackage{graphicx}
\usepackage{cite}
\usepackage{booktabs}
\usepackage[T1]{fontenc}
\usepackage{newtxmath}

\begin{document}

\title{SafeLink-Agent: Agentic Maintenance for Adaptive Bitrate Controllers over Dynamic Starlink Networks}
\author{Hongjun Xie, Bowen Zhang, Genke Yang, and Pengcheng Luo\thanks{This work was supported by the National Science and Technology Major Project of China under Grant 2025ZD1602400. \emph{(Corresponding author: Pengcheng Luo.)}}
\thanks{Hongjun Xie, Bowen Zhang, Genke Yang, and Pengcheng Luo are with Ningbo Artificial Intelligence Institute, Shanghai Jiao Tong University, Ningbo 315000, China, and also with the School of Automation and Intelligent Sensing, Shanghai Jiao Tong University, Shanghai 200240, China, and the Key Laboratory of System Control and Information Processing, Ministry of Education of China, Shanghai 200240, China (e-mail: xiehongjun@sjtu.edu.cn, bwz96sco@sjtu.edu.cn, gkyang@sjtu.edu.cn, luopeng69131@sjtu.edu.cn).}
\thanks{Hongjun Xie and Pengcheng Luo are also with Shanghai i-Space Orbital Computing Infrastructure Technology Co., Ltd., Shanghai 200235, China.}
\thanks{The source code will be available at: \protect\url{https://github.com/luopeng69131/SafeLink-Agent}.}
}

\maketitle

\begin{abstract}
Low Earth orbit (LEO) satellite broadband, represented by Starlink, is making high-resolution video streaming feasible beyond fixed terrestrial coverage. However, Starlink access links change across time and regions, exposing adaptive bitrate (ABR) streaming to shifting throughput tails, latency, volatility, and handover conditions. Existing ABR controllers are usually designed, tuned, or trained for specific network conditions, making it difficult to handle newly exposed hard Starlink profiles. This paper proposes \emph{SafeLink-Agent}, an agentic maintenance framework for ABR controllers over dynamic Starlink networks. SafeLink-Agent summarizes exposed failures and uses a large language model (LLM)-based agentic patch proposer to generate candidate patches, while replay verification determines whether each patch can be safely committed. The framework supports both rule-based controllers and learned controllers under the same maintenance workflow. Experiments on real Starlink networks show that SafeLink-Agent reduces the severe-session ratio of RobustMPC from 2.60\% to 0.40\% and reduces cumulative severe sessions from 45 to 7 in rolling maintenance. For learned controllers, verified adaptive auditing lowers the average severe-session ratio from 39.01\% to 9.79\%. These results demonstrate that agentic maintenance can improve ABR robustness under dynamic Starlink access conditions.

\end{abstract}


\section{Introduction}
\subsection{Background and Motivation}
\IEEEPARstart{S}{tarlink} and other low Earth orbit (LEO) satellite broadband systems are making high-bitrate video streaming feasible in rural, maritime, airborne, emergency, and other scenarios where terrestrial broadband is unavailable~\cite{michel2022first,kassem2022browser}. Beyond basic connectivity, large-scale LEO systems are also being developed to support low-latency computing and transmission services in remote areas~\cite{guo2025realtimeleo}. In such services, adaptive bitrate (ABR) control is the client-side mechanism that selects each video chunk bitrate according to the observed network condition, playback buffer, and chunk-size information. A higher bitrate improves visual quality when the link remains strong, but it can quickly drain the buffer and cause rebuffering when the available capacity drops.

This bitrate--buffer tradeoff becomes harder over Starlink because the access link is not a fixed terrestrial path. LEO constellations exhibit time-varying topologies and rapidly changing access conditions under high satellite mobility~\cite{han2023timevarying,li2026blockcyclic}. Their service continuity also depends on frequent handovers as satellites and users move across coverage regions~\cite{wang2025satelliteaircraft,huang2026dualthreshold}. In operational Starlink networks, satellite mobility, serving-satellite handovers, obstruction, regional coverage differences, traffic load, and gateway association can therefore change the throughput tail, volatility, and latency experienced by users~\cite{garcia2023multi,liu2025starnet}. As these conditions change after deployment, an ABR controller may encounter Starlink profiles that differ substantially from the conditions used for its original design or training. A configuration or policy that is suitable under stable access conditions may produce buffer depletion and severe session-level stalls under low-tail, high-volatility, high-latency, or handover-heavy Starlink profiles.

Most existing ABR studies still follow a static-controller design paradigm. Rule-based controllers such as BOLA and RobustMPC encode expert-designed control rules and fixed parameter settings~\cite{yin2015control,spiteri2020bola}, while learned controllers such as Pensieve and Comyco fix their policy parameters after training~\cite{mao2017neural,huang2019comyco}. Recent learned ABR methods further improve adaptation and generalization through curriculum learning, offline reinforcement learning, meta-learning, and bitrate guidance~\cite{xia2022genet,yi2025optimizing,bentaleb2023meta,bentaleb2024bitrate,li2023metaabr,kan2025merina+}. However, once deployed, both rule-based and learned controllers are difficult to adjust when newly observed Starlink profiles expose new failure modes: rule-based controllers often require manual redesign or retuning, and learned controllers often require additional data collection and retraining.

\begin{figure*}[!t]
  \centering
  \includegraphics[width=0.88\textwidth]{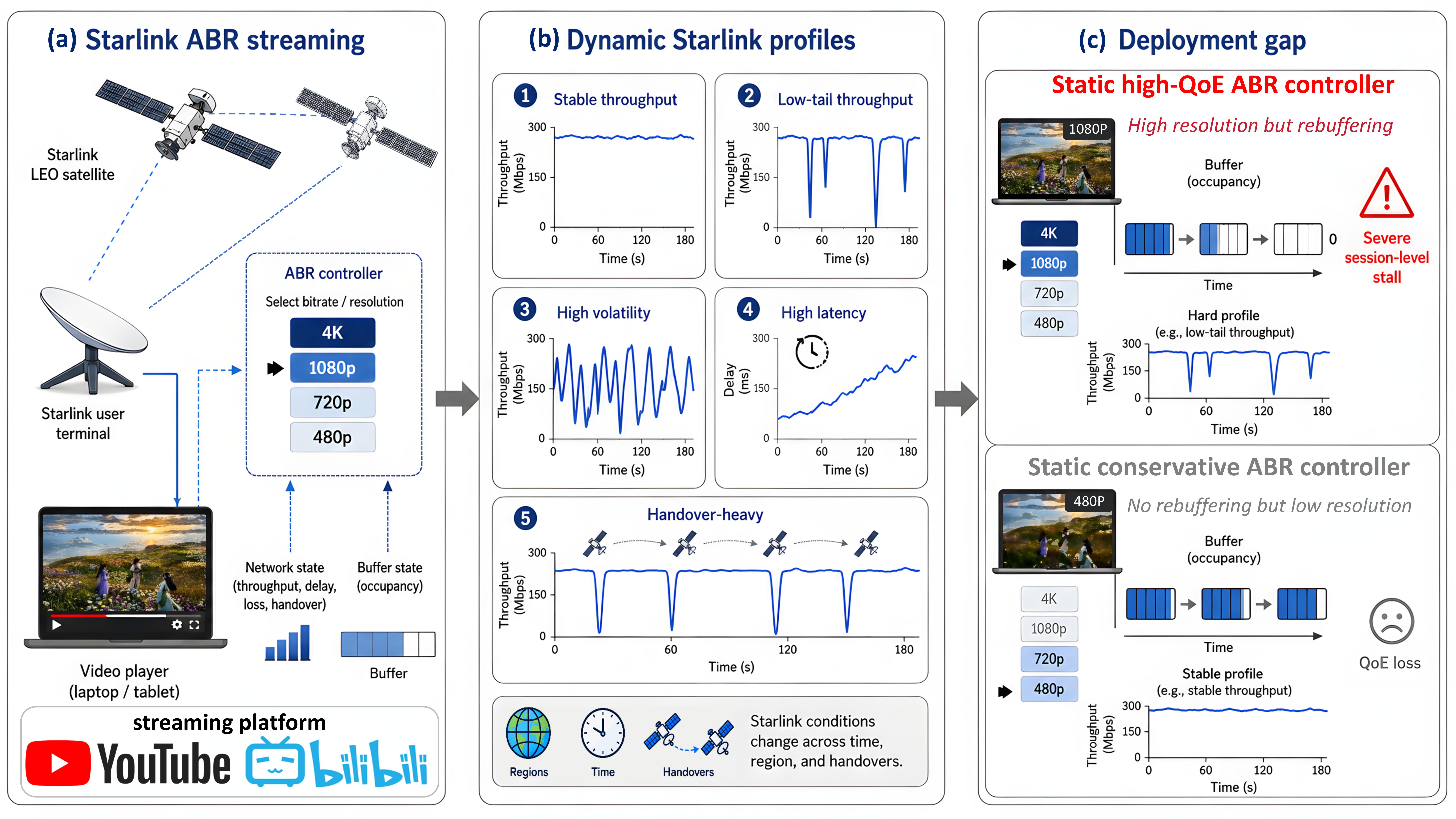}
  \caption{Deployment gap of ABR streaming over dynamic Starlink access links. (a) A Starlink video client selects chunk bitrate or resolution according to network and buffer states. (b) Starlink access conditions vary across regions, time, latency, and handovers, producing stable, low-tail, high-volatility, high-latency, and handover-heavy profiles. (c) A Static high-QoE ABR controller can drain the playback buffer and cause severe session-level stalls on hard profiles, whereas Static conservative control avoids rebuffering by sacrificing resolution on stable profiles.}
  \label{fig:intro-dynamic-gap}
\end{figure*}

This deployment gap leaves two unsatisfactory choices. Keeping the originally designed or trained controller as a Static default may preserve high average Quality of Experience (QoE) on common profiles, but it can leave severe stalls on newly exposed hard profiles. Applying a Static conservative configuration or safety rule can reduce such stalls, but it may unnecessarily lower video resolution on stable profiles where the original controller already works well. Fig.~\ref{fig:intro-dynamic-gap} illustrates the Starlink ABR scenario and the resulting deployment gap. Therefore, dynamic Starlink ABR requires a maintenance view: the controller should be locally repaired for profiles where it fails, while regression checks should prevent the repair from damaging profiles where the controller already behaves well.

Such a maintenance view requires the ability to read replay evidence, reason about failure causes, propose localized controller changes, and use verification feedback. This requirement matches the emerging strengths of large language models (LLMs) and coding agents, whose reasoning, repository-level code editing, and tool-use capabilities have been demonstrated in general reasoning models, networking tasks, and software-engineering agents~\cite{openai2023gpt4,openai2024o1,wu2024netllm,yang2024sweagent,zhang2024autocoderover}. These capabilities make LLMs and coding agents suitable for maintenance tasks that involve diagnosis, repair proposal, self-improvement, and auditing, and allow their outputs to be organized as structured and auditable maintenance candidates~\cite{jimenez2024swebench,xia2025agentless,lascasas2024llexus,weng2026learning_beyond_gradients}.

\begin{figure}[!t]
  \centering
  \includegraphics[width=\columnwidth]{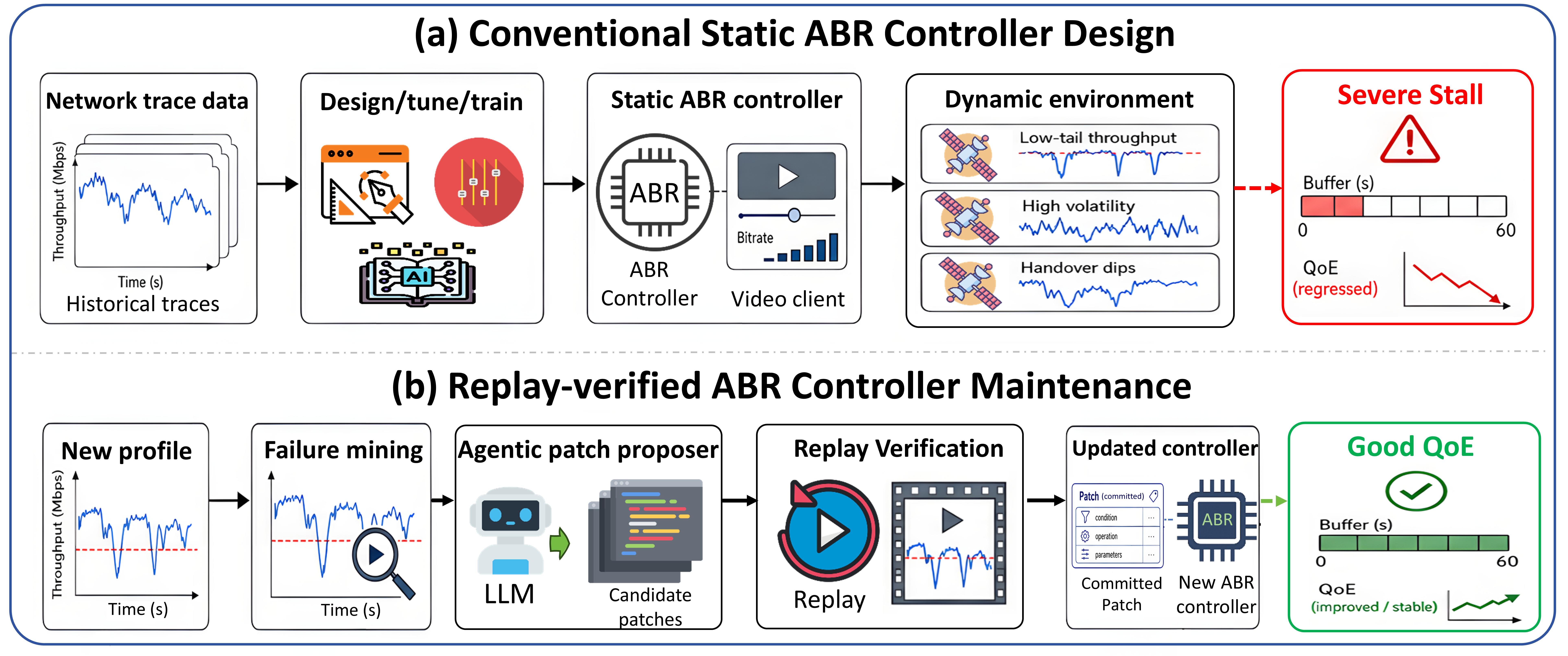}
  \caption{From conventional ABR controller design to agentic controller maintenance. (a) Conventional ABR design builds a static controller from historical trace data, but future Starlink profiles may expose severe stalls and QoE regression. (b) SafeLink-Agent treats new profiles as maintenance evidence: it diagnoses replay failures, uses an LLM-based agentic patch proposer to generate candidate patches, and updates the controller only after replay verification.}
  \label{fig:design-to-maintenance}
\end{figure}

Motivated by this observation, this paper proposes \emph{SafeLink-Agent}, an agentic maintenance framework for ABR controllers over dynamic Starlink networks. SafeLink-Agent mines Starlink profiles and ABR replay logs, summarizes newly exposed failure patterns, uses an agentic patch proposer to generate a candidate patch, and commits the patch only after target-profile replay and regression-memory replay confirm that it reduces severe-session risk without introducing unacceptable QoE regression. Fig.~\ref{fig:design-to-maintenance} summarizes this shift from static-controller design to agentic controller maintenance.

\subsection{Contributions}
The main contributions of this paper are summarized as follows:
\begin{itemize} 
\item We formulate dynamic ABR controller maintenance over Starlink networks. Instead of treating ABR as a static-controller design problem, the formulation models controller updates, Starlink profiles, candidate patches, severe-session risk metrics, and verifier constraints.

\item We propose \emph{SafeLink-Agent}, an agentic maintenance framework for ABR controllers over dynamic Starlink networks. SafeLink-Agent mines profile and failure evidence from Starlink replay logs, uses an LLM-based agentic patch proposer to generate candidate patches, and commits a patch only after replay verification confirms target-profile improvement without unacceptable regression.

\item We instantiate SafeLink-Agent for both rule-based and learned ABR controllers. For BOLA and RobustMPC, it produces profile-conditioned configuration patches; for SABR, Pensieve, and Comyco, it produces verified runtime-auditing patches. This demonstrates that heterogeneous ABR controllers can be maintained through a unified patch-and-verify interface.

\item We build a dynamic Starlink ABR evaluation workflow from real Starlink traces. The experiments evaluate SafeLink-Agent across diverse controller types and newly observed Starlink profile scenarios. The results show that agentic maintenance can reduce severe-session failures while avoiding QoE loss.
\end{itemize}

\subsection{Organization}
The remainder of this paper is organized as follows. Section II reviews related work on ABR control under dynamic network conditions and LLM-based system maintenance. Section III formulates dynamic Starlink ABR as a controller-maintenance problem. Section IV presents the SafeLink-Agent maintenance framework, including profile mining, candidate patch generation, patch schema validation, and patch commitment. Section V describes the experimental setting and evaluates SafeLink-Agent across newly observed Starlink profiles. Finally, Section VI concludes the paper.

\section{Related Work}

The related literature is organized along two lines. We first review ABR control under dynamic network conditions, which is the application domain of this paper. We then discuss LLM agents for system maintenance and program repair, which motivate the agentic maintenance of SafeLink-Agent.

\subsection{ABR Control under Dynamic Network Conditions}

ABR control has long been studied as the client-side mechanism for adapting video quality to network dynamics. Classical online methods select chunk bitrates using throughput prediction, buffer occupancy, or short-horizon optimization. Model Predictive Control (MPC)-style ABR explicitly plans bitrate decisions over a future horizon~\cite{yin2015control}, while BOLA formulates bitrate selection as a buffer-based utility maximization problem~\cite{spiteri2020bola}. Prediction-assisted systems such as Lumos and Starnet further show that more accurate throughput prediction can improve mobile video QoE~\cite{lv2022lumos,lv2024accurate,liu2025starnet}. These methods can react to short-term network variation, but their control rules and parameters are usually fixed once deployed.

Learning-based ABR controllers improve this design space by learning policies from QoE feedback. Pensieve trains a neural ABR policy with reinforcement learning~\cite{mao2017neural}, and Comyco combines imitation learning and reinforcement learning to improve ABR decisions~\cite{huang2019comyco}. SABR further follows a pretraining and fine-tuning design for satellite ABR~\cite{luo2025sabr}. More recent studies focus on adaptation and generalization across heterogeneous traces. Genet uses automatic curriculum generation to expose a policy to diverse network conditions~\cite{xia2022genet}. Offline reinforcement learning and meta-learning improve ABR policies across traces or tasks~\cite{yi2025optimizing,bentaleb2023meta}, while bitrate guidance, MetaABR, and Merina+ explore additional mechanisms for cross-condition adaptation~\cite{bentaleb2024bitrate,li2023metaabr,kan2025merina+}.

Starlink access makes this post-deployment question more important because measured links show temporal, regional, and handover-related performance variation~\cite{garcia2023multi,liu2025starnet}. Video studies over Starlink and LEO networks further indicate that such variation can affect real-time and streaming applications~\cite{zhao2023realtime,zhao2024dash}. However, existing ABR research mainly asks how to design, train, or generalize a controller before deployment. Even when a controller adapts online at the chunk level, its rule set, policy, or safety mechanism is typically not repaired after new hard profiles are discovered.

These limitations leave a gap between ABR controller design and controller maintenance under dynamic Starlink access. Existing ABR studies mainly optimize the controller before deployment or adapt bitrate decisions online at the chunk level, but they rarely study how an already deployed controller should be diagnosed, repaired, and regression-tested when newly observed profiles expose severe stalls or QoE regression. This gap motivates treating ABR controllers as maintainable artifacts after deployment, rather than as fixed outputs of an offline design process.

\subsection{LLM Agents for System Maintenance and Program Repair}

Recent progress in LLM-based software engineering shows that language models can operate on real repositories rather than isolated code snippets. SWE-bench evaluates models on GitHub issue resolution tasks that require repository-level understanding and patch generation~\cite{jimenez2024swebench}. SWE-agent further demonstrates that agent-computer interfaces, repository navigation, file editing, and test execution can substantially improve automated software engineering performance~\cite{yang2024sweagent}. AutoCodeRover combines LLMs with program-structure-aware code search and test feedback for autonomous program improvement~\cite{zhang2024autocoderover}. Meanwhile, Agentless shows that careful localization, localized repair, and patch validation can be as important as complex agent scaffolding~\cite{xia2025agentless}. These studies support the use of LLMs as maintenance assistants, but they also emphasize that generated patches require external validation.

LLM agents have also begun to appear in system operation and networking contexts. LLexus uses LLM agents to convert troubleshooting guides into executable incident-management plans~\cite{lascasas2024llexus}. LogGPT explores generative pre-trained transformer (GPT)-based log anomaly detection for system diagnosis~\cite{han2023loggpt}. NetLLM adapts LLMs for networking tasks and shows the potential of LLM reasoning in networked systems~\cite{wu2024netllm}. These works are relevant because ABR controller maintenance also involves logs, configuration choices, and network-dependent behavior. However, a latency-sensitive ABR controller cannot simply trust a generated action or patch because the consequence is measured in playback stalls and QoE regression.

These works indicate that LLM and coding agents can support the maintenance of deployed systems, but the validation target in ABR control differs from those in existing settings. In software repair, correctness is commonly evaluated by functional tests; in artificial intelligence for IT operations (AIOps), the output is often assessed as a diagnosis or execution plan. For an ABR controller, a maintenance proposal affects playback behavior over measured network traces: a change that mitigates failures on a newly observed profile may also reduce QoE or introduce stalls on profiles that were previously stable. Therefore, LLM-based agentic ABR maintenance requires a way to generate constrained controller changes and judge their effect on both newly exposed failures and previously acceptable playback behavior.

\section{Dynamic ABR Controller Maintenance Formulation}
\label{sec:formulation}

This section formulates ABR control over dynamic Starlink access as a maintenance problem. We first define Starlink profiles as the operating units on which new failures are observed. We then define replay-based evaluation to measure how a controller behaves on each profile. With the measurement interface specified, we introduce the controller update to describe what SafeLink-Agent can change. Finally, we define the verifier constraints that determine whether a candidate patch should be committed.

\subsection{Starlink Profiles}
Starlink measurements are first converted into replay traces, each of which represents one streaming session under a measured throughput and latency trajectory. Controller maintenance, however, should not react to an isolated trace alone; it should react when a group of traces exposes a recurring operating condition or failure mode. We call such a group a \emph{Starlink profile}. A profile may correspond to a deployment stage, such as a newly collected calibration batch, or to a hard access condition, such as low throughput tail, high volatility, high latency, or frequent satellite changes. 

As new profiles are exposed to the deployed system, the access environment is represented as a sequence
\begin{equation}
    D_1,D_2,\ldots,D_k,\ldots ,
    \label{eq:dynamic-profiles}
\end{equation}
where $D_k$ denotes the $k$th profile exposed to the maintenance workflow. Formally, $D_k$ is a set of measured replay sessions:
\begin{equation}
    D_k=\{\tau_{k,1},\tau_{k,2},\ldots,\tau_{k,n_k}\},
    \label{eq:profile-trace-set}
\end{equation}
where each $\tau_{k,j}$ is one replay trace in this profile and $n_k$ is the number of sessions. To characterize the profile for maintenance, we summarize $D_k$ by metadata
\begin{equation}
    z(D_k)=
    \left[
    \mu_{\mathrm{tp}},
    q_{0.05}^{\mathrm{tp}},
    \sigma_{\mathrm{tp}},
    \mu_{\mathrm{lat}},
    H_{\mathrm{ho}}
    \right]^{\top},
    \label{eq:profile-summary}
\end{equation}
where $z(D_k)$ is the compact description of profile $D_k$, $\mu_{\mathrm{tp}}$ is the mean throughput, $q_{0.05}^{\mathrm{tp}}$ is the 5th-percentile throughput, $\sigma_{\mathrm{tp}}$ is throughput volatility, $\mu_{\mathrm{lat}}$ is mean latency, and $H_{\mathrm{ho}}$ represents handover or serving-satellite change intensity. These metadata describe the environment in which a controller is replayed and help SafeLink-Agent distinguish, for example, a low-tail profile from a high-volatility or handover-heavy profile.

The profile sequence also defines what should be protected during maintenance. When the current profile is $D_k$, the regression memory is formed from previously validated profiles and representative sessions:
\begin{equation}
    \mathcal{M}_k
    =
    \{D_i: i<k,\; D_i\ \text{is validated}\}
    \cup
    \mathcal{S}_{\mathrm{mem}},
    \label{eq:regression-memory}
\end{equation}
where $\mathcal{S}_{\mathrm{mem}}$ denotes representative sessions retained from earlier replays. The regression memory stores previously acceptable playback behavior, so that later maintenance decisions can be checked against profiles and sessions that should remain stable.

\subsection{Replay-Based Controller Evaluation}
\label{sec:replay-based-evaluation}
After the profile sequence is defined, the next step is to specify how a controller is evaluated on each profile. Controller maintenance requires a behavioral interface that can compare the original and patched controllers under the same measured network conditions. Replay evaluation provides this interface: it exposes not only aggregate QoE, but also the logs needed for failure mining and regression checking. Given a controller $C$ and a measured Starlink streaming session $\tau$, the replay simulator executes the controller over the measured throughput sequence and returns session-level outcomes:
\begin{equation}
    \left(Q(C,\tau),B(C,\tau),\ell(C,\tau)\right)
    =
    \operatorname{Replay}(C,\tau),
    \label{eq:replay-interface}
\end{equation}
where $Q(C,\tau)$ is the session-level QoE, $B(C,\tau)$ is the cumulative rebuffering, and $\ell(C,\tau)$ is the replay log that records actions, buffer states, rebuffering events, and other diagnostic signals. This replay interface is the basic object maintained by SafeLink-Agent: a patch is judged by how the patched controller behaves under trace replay.

The QoE used in replay follows the standard ABR form that combines video quality, rebuffering penalty, and smoothness penalty. For a session with $T$ chunks, we write it as
\begin{equation}
\begin{aligned}
Q(C,\tau)
&=
\sum_{t=1}^{T}
\Bigl[
u(a_t)-\mu\rho_t  \\
&\qquad
-\eta |u(a_t)-u(a_{t-1})|
\Bigr],\\
B(C,\tau)
&=
\sum_{t=1}^{T}\rho_t .
\end{aligned}
    \label{eq:replay-qoe-rebuf}
\end{equation}
Here $u(a_t)$ is the utility of the selected bitrate action, $\rho_t$ is the rebuffering duration of chunk $t$, and $\mu$ and $\eta$ are the rebuffering and smoothness penalty weights. The exact simulator computes $\rho_t$ from the measured throughput, chunk size, and buffer evolution.

For a profile $D$, replay outcomes are aggregated over the sessions in that profile:
\begin{equation}
\begin{aligned}
\bar Q(C,D)
&=
\mathbb{E}_{\tau\sim D}[Q(C,\tau)],\\
\bar B(C,D)
&=
\mathbb{E}_{\tau\sim D}[B(C,\tau)] .
\end{aligned}
    \label{eq:profile-replay-aggregate}
\end{equation}
These profile-level quantities provide the playback measurements used by later maintenance decisions. In particular, they allow SafeLink-Agent to compare controller behavior on both the newly exposed profile and the regression memory.

\subsection{Controller Update}
Profiles and replay evaluation define where controller behavior is observed and how playback outcomes are measured. The maintenance formulation also needs to specify how an already deployed controller may be updated. We model this update as a versioned patch operation: the current controller remains the reference controller, and any proposed change must belong to an admissible patch space.

Let $C_v$ denote the current deployed ABR controller version. The controller can be rule-based, learned, or risk-aware. When a new Starlink profile exposes severe-session failures, SafeLink-Agent searches for a candidate patch from the admissible patch space and applies it to obtain the next controller version:
\begin{equation}
    C_{v+1}=C_v\oplus p,\quad p\in\mathcal{P}.
    \label{eq:controller-patch}
\end{equation}
Here $\oplus$ denotes applying a patch to the current controller. A patch is the maintenance decision made for the controller version. The patch $p$ may change a finite configuration, add a profile-conditioned rule, enable a runtime auditor, introduce a fallback condition, or update a constrained training configuration. Its complexity is denoted by $\Omega(p)$ and is limited to keep the repair interpretable and verifiable.

\subsection{Risk Metrics and Verifier Constraints}
After a controller update is represented as a patch operation, SafeLink-Agent needs criteria for deciding whether the patched controller is acceptable. The verifier first measures severe-session risk and worst-tail rebuffering on replayed profiles, and then checks whether the patch improves the newly exposed profile without causing unacceptable regression.

We first define the severe-session ratio of controller $C$ on profile $D$ as
\begin{equation}
    R_{\mathrm{sev}}(C,D)
    =
    \mathbb{E}_{\tau\sim D}
    \left[
    \mathbf{1}\{B(C,\tau)>b_{\mathrm{sev}}\}
    \right],
    \label{eq:severe-risk-maintenance}
\end{equation}
where $\mathbf{1}\{\cdot\}$ is the indicator function and $b_{\mathrm{sev}}$ is the severe-stall threshold. We further define the worst-tail rebuffering metric
\begin{equation}
    T_{\alpha}(C,D)
    =
    \mathbb{E}_{\tau\in \operatorname{Tail}_{\alpha}(C,D)}
    \left[
    B(C,\tau)
    \right],
    \label{eq:tail-risk-maintenance}
\end{equation}
where $\operatorname{Tail}_{\alpha}(C,D)$ denotes the worst $\alpha$ fraction of sessions in profile $D$ ranked by cumulative rebuffering under controller $C$. 

These replay metrics become the quantities checked by the verifier. On the newly exposed profile, the patched controller should repair the observed failure without an excessive QoE cost. Because such a local repair may also affect profiles that were previously acceptable, the verifier checks the same patch on regression memory and requires the preserved behavior to remain within tolerance. Accordingly, for a current controller $C_v$, a newly exposed profile $D_{\mathrm{new}}$, and regression memory $\mathcal{M}$, we collect the acceptance requirements as the verifier constraints $\Gamma$:
\begin{equation}
\begin{aligned}
\Gamma:\quad
& p\in\mathcal{P}, \\
& R_{\mathrm{sev}}(C_v\oplus p,D_{\mathrm{new}})
   \leq R_{\mathrm{sev}}(C_v,D_{\mathrm{new}})-\delta, \\
& T_{\alpha}(C_v\oplus p,D_{\mathrm{new}})
   \leq T_{\alpha}(C_v,D_{\mathrm{new}})-\gamma, \\
& \bar Q(C_v\oplus p,D_{\mathrm{new}})
   \geq \bar Q(C_v,D_{\mathrm{new}})-\epsilon, \\
& R_{\mathrm{sev}}(C_v\oplus p,\mathcal{M})
   \leq R_{\mathrm{sev}}(C_v,\mathcal{M})+\zeta, \\
& \bar Q(C_v\oplus p,\mathcal{M})
   \geq \bar Q(C_v,\mathcal{M})-\epsilon_{\mathrm{reg}}, \\
& \Omega(p)\leq \kappa .
\end{aligned}
\label{eq:verifier-constraints}
\end{equation}
The first two inequalities require the patch to reduce severe-session risk and worst-tail rebuffering on the target profile. The third inequality limits the QoE cost on that profile. The fourth and fifth inequalities implement regression memory: a patch should not introduce new severe-session risk or excessive QoE loss on previously stable profiles. The final inequality limits patch complexity. This formulation separates proposal generation from patch acceptance: the proposer may generate a candidate patch, but the patch is committed only if replay verification confirms that $p$ satisfies $\Gamma$.

\section{SafeLink-Agent Maintenance}
\label{sec:safelink-agent}

\begin{figure*}[!t]
  \centering
  \includegraphics[width=0.98\textwidth]{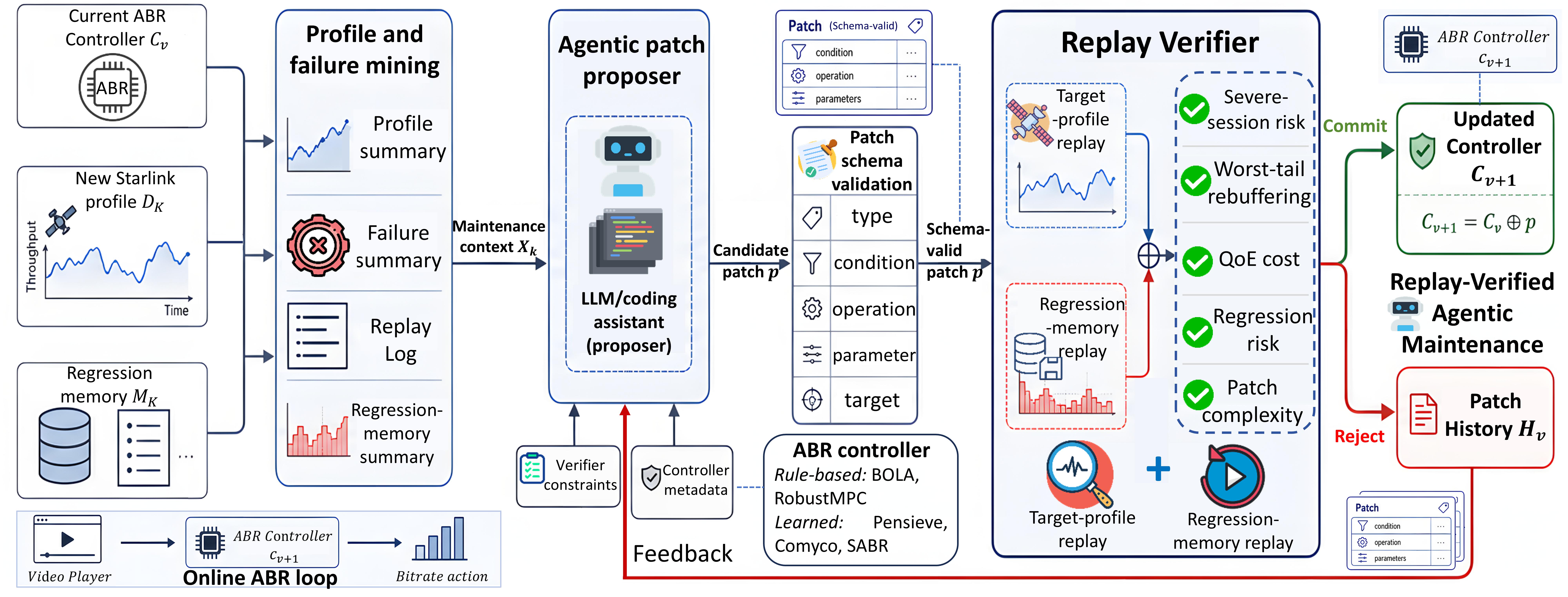}
  \caption{Overall framework of SafeLink-Agent. The current ABR controller, a newly observed Starlink profile, and regression memory are first summarized by profile and failure mining. The agentic patch proposer generates a candidate patch for rule-based or learned ABR controllers, but patch commitment is decided by the replay verifier using target-profile replay, regression-memory replay, QoE cost, regression risk, and patch-complexity checks. The online ABR loop remains separate from the agentic maintenance workflow.}
  \label{fig:safelink-framework}
\end{figure*}

SafeLink-Agent realizes controller maintenance as a sequence of failure summarization, candidate patch generation, and verified patch commitment. When a newly observed Starlink profile exposes a maintenance trigger, the framework first summarizes where the current controller fails. Using this summary, the agentic patch proposer then generates a candidate patch. Patch schema validation checks whether the patch follows the required patch schema, and replay verification finally decides whether it should be committed to produce the new controller. Fig.~\ref{fig:safelink-framework} summarizes this maintenance workflow.

\subsection{Profile and Failure Mining}
Profile and failure mining converts replay evidence into the structured maintenance context used by the agentic patch proposer. It has two roles. First, it checks whether the newly observed profile $D_k$ actually requires maintenance. Second, if maintenance is triggered, it summarizes the Starlink profile, the observed playback failure, representative hard sessions, and the behavior that should be preserved.

The trigger check is performed after the current controller $C_v$ is replayed on $D_k$ using the interface in Section~\ref{sec:replay-based-evaluation}. The regression memory $\mathcal{M}_k$ provides the reference behavior for identifying unacceptable regression. For the two quantitative risk triggers, we define
\begin{equation}
\mathcal{I}_{\mathrm{trig}}(C_v,D_k)
=
\mathbf{1}\!\left[
R_{\mathrm{sev}}(C_v,D_k)>\beta_{\mathrm{sev}}
\vee
T_{\alpha}(C_v,D_k)>\beta_{\mathrm{tail}}
\right],
\label{eq:maintenance-trigger}
\end{equation}
where $\beta_{\mathrm{sev}}$ and $\beta_{\mathrm{tail}}$ are the trigger thresholds for severe-session ratio and worst-tail rebuffering, respectively. Visible QoE regression relative to regression memory can be used as auxiliary trigger evidence. When $\mathcal{I}_{\mathrm{trig}}(C_v,D_k)=1$, the miner builds the maintenance context described below.

The context first records what kind of Starlink profile is being handled. For the current profile $D_k$, the profile-side summary is the metadata vector $z(D_k)$ defined in \eqref{eq:profile-summary}. It records network-side properties such as throughput tail, volatility, latency, and handover intensity, which distinguish, for example, a low-tail profile from a high-volatility or handover-heavy profile.

The second part of the context describes how the current controller behaves on this profile. The miner uses the replay outputs of $C_v$ on $D_k$ to form an aggregate failure summary:
\begin{equation}
s(C_v,D_k)=
\left[
\bar Q,
\bar B,
R_{\mathrm{sev}},
T_{\alpha},
N_{\mathrm{sev}},
\rho_{\mathrm{audit}}
\right]^{\top},
\label{eq:failure-summary}
\end{equation}
where $N_{\mathrm{sev}}$ is the number of severe sessions and $\rho_{\mathrm{audit}}$ is the audit-intervention rate when applicable. This summary gives the agentic patch proposer aggregate evidence of QoE, rebuffering, tail risk, and auditing behavior.

Aggregate metrics alone do not show which sessions caused the failure. The miner therefore uses the detailed replay log $\ell(C_v,\tau)$ to identify representative hard sessions:
\begin{equation}
\begin{aligned}
\mathcal{F}_{\mathrm{sev}}(C_v,D_k)
&=
\{\tau\in D_k\mid B(C_v,\tau)>b_{\mathrm{sev}}\},\\
\mathcal{F}_{\mathrm{tail}}(C_v,D_k)
&=
\operatorname{Tail}_{\alpha}(C_v,D_k).
\end{aligned}
\label{eq:failure-sets}
\end{equation}
$\mathcal{F}_{\mathrm{sev}}$ contains sessions whose cumulative rebuffering exceeds the severe-stall threshold, and $\mathcal{F}_{\mathrm{tail}}$ contains the worst-tail sessions ranked by cumulative rebuffering. These sets connect the profile-level failure to concrete replay sessions that can be inspected by the agentic patch proposer and later checked by the verifier.

Finally, the profile summary, aggregate failure summary, representative hard sessions, and regression-memory summary are combined into the maintenance context:
\begin{equation}
\mathcal{X}_k
=
\left(
z(D_k),
s(C_v,D_k),
\mathcal{F}_{\mathrm{sev}},
\mathcal{F}_{\mathrm{tail}},
s(C_v,\mathcal{M}_k)
\right),
\label{eq:maintenance-context}
\end{equation}
where $s(C_v,\mathcal{M}_k)$ summarizes the playback behavior recorded in regression memory. Thus, $\mathcal{X}_k$ tells the agentic patch proposer what kind of profile is being handled, how the current controller fails on that profile, which sessions represent the failure, and what behavior should be preserved.

\subsection{Agentic Patch Proposer}

The agentic patch proposer uses an LLM assistant to generate a candidate repair for the ABR controller. After profile and failure mining, it maps the structured maintenance inputs to a candidate patch:
\begin{equation}
    p =
    G_{\theta}
    \left(
    \mathcal{X}_k,
    \mathcal{H}_v,
    \mathrm{meta}(C_v),
    \Gamma
    \right),
    \label{eq:agent-proposal-map}
\end{equation}
where $G_{\theta}$ denotes the proposer, which can be implemented by an LLM assistant; $\mathcal{H}_v$ denotes Patch History for controller version $v$; and $\mathrm{meta}(C_v)$ denotes the controller metadata.

The controller metadata gives the proposer controller-specific information before a patch is generated. It describes which parameters of the current controller are maintainable and which profile or failure features can be used to condition a maintenance action. Table~\ref{tab:controller-metadata} presents the detailed controller metadata settings used in this paper.

\begin{table*}[!t]
\centering
\caption{Controller metadata settings used by SafeLink-Agent.}
\label{tab:controller-metadata}
\footnotesize
\begin{tabular}{@{}p{0.15\textwidth}p{0.36\textwidth}p{0.41\textwidth}@{}}
\toprule
\textbf{Controller} & \textbf{Maintainable Parameters} & \textbf{Available Profile/Failure Features} \\
\midrule
BOLA & Buffer-threshold candidates, e.g., $(10,30)$ s or $(15,45)$ s. & Throughput tail, volatility, buffer pressure, and severe-session evidence. \\
RobustMPC & Prediction horizon, safety margin, and bitrate-cap candidates, e.g., horizon $5/6$ and margin $1.00/0.85$. & Tail throughput, latency, volatility, handover intensity, and replay failure summaries. \\
Learned ABR & Policy identity and admissible runtime auditor parameters such as safe-capacity margin, low-buffer threshold, and buffer guard. & Buffer state, safe capacity, requested bitrate risk, and severe-session examples. \\
\bottomrule
\end{tabular}
\end{table*}

The output $p$ is represented as a structured candidate patch rather than free-form code. It specifies the target controller version, patch type, activation condition, maintenance operation, and bounded parameter values. The operation forms considered in this paper are configuration selection, profile-conditioned configuration rules, and runtime auditor attachment or tuning for learned ABR policies. 

\subsection{Patch Schema Validation}

The agentic proposer may return an unsupported patch type, unknown fields, invalid activation conditions, or parameter values outside the allowed range. A candidate patch is schema-valid only if it conforms to this controller-specific schema and satisfies the patch-complexity bound. 

The admissible patch space can therefore be written as
\begin{equation}
\mathcal{P}
=
\{p\mid \operatorname{Schema}_{C_v}(p)=1,\;
\Omega(p)\leq\kappa\}.
\label{eq:typed-patch-space}
\end{equation}
Here, $\operatorname{Schema}_{C_v}(p)=1$ means that the candidate patch matches the patch schema selected for $C_v$, including the controller type, allowed operation, approved activation-condition fields, and admissible parameter ranges. 

If a candidate patch satisfies \eqref{eq:typed-patch-space}, patch schema validation outputs it as a \emph{schema-valid patch} and sends it to replay verification. If the candidate patch fails validation, the module outputs a schema-violation record and stores it in Patch History before replay begins.

\subsection{Replay Verification and Patch Commit}

\begin{figure}[!t]
  \centering
  \includegraphics[width=\columnwidth]{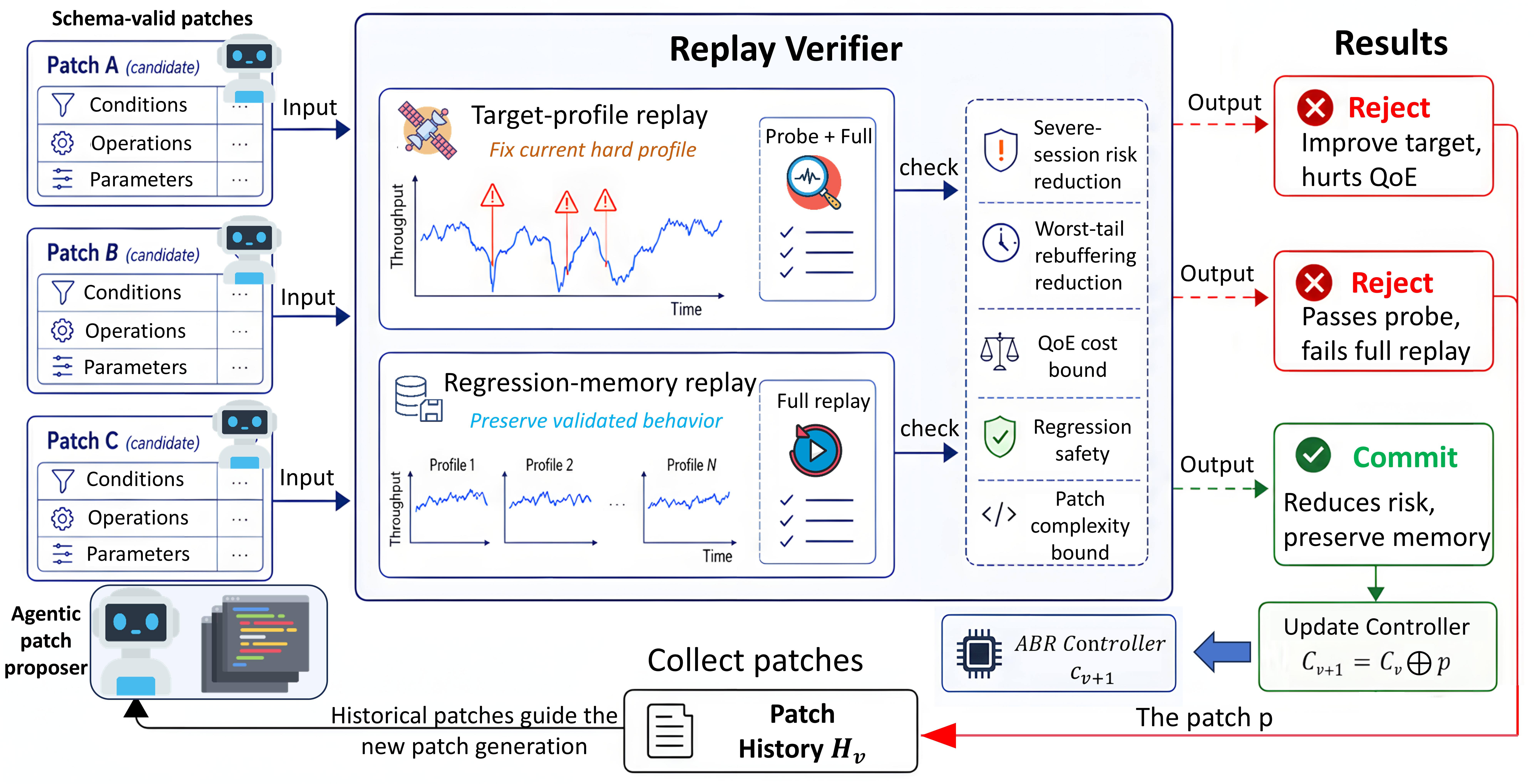}
  \caption{Replay verification for schema-valid patches. The verifier evaluates each patch on the challenging target profile and on previously validated profiles in regression memory.}
  \label{fig:replay-verifier}
\end{figure}

Replay verification is the commit gate between a schema-valid patch and a new controller version. The verification is organized into two checks: target-profile replay examines whether the patch repairs the newly exposed failure, and regression-memory replay examines whether it preserves previously validated behavior. It reuses the replay-based evaluation in Section~\ref{sec:replay-based-evaluation} to compare the current controller and the patched controller under measured Starlink profiles. Fig.~\ref{fig:replay-verifier} summarizes the verification flow.

\subsubsection{Target-Profile Replay}
Target-profile replay checks whether a candidate patch actually addresses the failure that triggered maintenance. For efficiency, a lightweight probe replay may first evaluate the patch on a small subset of $D_k$ to screen clearly ineffective proposals. Passing this probe does not certify the patch and does not check regression memory; it only sends the patch to full target-profile replay.

In the full replay, the current controller $C_v$ and the patched controller $C_v\oplus p$ are evaluated on the same target profile $D_k$, so any behavioral difference can be attributed to the patch rather than to a change in traces. The verifier then checks the target-profile block of $\Gamma$, obtained from \eqref{eq:verifier-constraints} by setting $D_{\mathrm{new}}=D_k$:
\begin{equation}
\begin{aligned}
\Gamma_{\mathrm{tar}}:\quad
& R_{\mathrm{sev}}(C_v\oplus p,D_k)
   \leq R_{\mathrm{sev}}(C_v,D_k)-\delta,\\
& T_{\alpha}(C_v\oplus p,D_k)
   \leq T_{\alpha}(C_v,D_k)-\gamma,\\
& \bar Q(C_v\oplus p,D_k)
   \geq \bar Q(C_v,D_k)-\epsilon .
\end{aligned}
\label{eq:target-constraints}
\end{equation}
Here, the first two inequalities require reductions in severe-session risk and worst-tail rebuffering on $D_k$, while the third limits the QoE cost of the repair. 

\subsubsection{Regression-Memory Replay}
Regression-memory replay evaluates the same candidate patch on profiles whose playback behavior should be preserved. This step distinguishes a local repair from a global retuning: a patch may reduce stalls on $D_k$ but still be rejected if it lowers QoE or increases severe-session risk on $\mathcal{M}$. The verifier therefore checks the regression-memory block of $\Gamma$:
\begin{equation}
\begin{aligned}
\Gamma_{\mathrm{mem}}:\quad
& R_{\mathrm{sev}}(C_v\oplus p,\mathcal{M})
   \leq R_{\mathrm{sev}}(C_v,\mathcal{M})+\zeta,\\
& \bar Q(C_v\oplus p,\mathcal{M})
   \geq \bar Q(C_v,\mathcal{M})-\epsilon_{\mathrm{reg}}.
\end{aligned}
\label{eq:memory-constraints}
\end{equation}
Here, the first inequality limits severe-session risk increase on profiles in regression memory, and the second limits QoE regression.

\subsubsection{Patch Commit and Patch History}
The verifier converts the target-profile replay and regression-memory replay reports into an accept-or-reject decision. A patch is accepted only when patch schema validation and these checks confirm that $p$ satisfies $\Gamma$. Rejected patches are stored in the Patch History $\mathcal{H}_v$, and a machine-readable reason, such as insufficient target-risk reduction, excessive QoE cost, regression-memory constraint violation, or patch-complexity violation.

The committed controller update is therefore
\begin{equation}
C_{v+1}
=
\begin{cases}
C_v\oplus p, & \text{if } p\ \text{satisfies}\ \Gamma,\\
C_v, & \text{otherwise}.
\end{cases}
\label{eq:commit-rule}
\end{equation}

Algorithm~\ref{alg:safelink-maintenance} summarizes the SafeLink-Agent maintenance procedure.

\begin{algorithm}[!t]
\caption{SafeLink-Agent Maintenance Procedure}
\label{alg:safelink-maintenance}
\begin{algorithmic}[1]
\STATE \textbf{Input:} current controller $C_v$, new profile $D_k$, regression memory $\mathcal{M}_k$, Patch History $\mathcal{H}_v$, controller metadata $\mathrm{meta}(C_v)$, verifier constraints $\Gamma$
\STATE Replay $C_v$ on $D_k$ using \eqref{eq:replay-interface}.
\STATE Compute the maintenance trigger $\mathcal{I}_{\mathrm{trig}}(C_v,D_k)$ by \eqref{eq:maintenance-trigger}.
\IF{$\mathcal{I}_{\mathrm{trig}}(C_v,D_k)=0$}
    \STATE Keep $C_{v+1}\leftarrow C_v$ and store $D_k$ as validated replay evidence.
    \STATE \textbf{return} $C_{v+1}$ and the replay report.
\ENDIF
\STATE Mine $z(D_k)$, $s(C_v,D_k)$, $\mathcal{F}_{\mathrm{sev}}$, and $\mathcal{F}_{\mathrm{tail}}$ from replay outputs.
\STATE Build maintenance context $\mathcal{X}_k$ using \eqref{eq:maintenance-context}.
\STATE Generate candidate patch $p$ with the agentic patch proposer using \eqref{eq:agent-proposal-map}.
\STATE Perform patch schema validation for $p$ using \eqref{eq:typed-patch-space}.
\IF{$p$ is not schema-valid}
    \STATE Store the schema-violation reason in Patch History $\mathcal{H}_v$.
    \STATE Keep $C_{v+1}\leftarrow C_v$.
\ELSE
    \STATE Run target-profile replay for $C_v\oplus p$ on $D_k$ and check $\Gamma_{\mathrm{tar}}$.
    \STATE Run regression-memory replay for $C_v\oplus p$ on $\mathcal{M}_k$ and check $\Gamma_{\mathrm{mem}}$.
    \IF{$p$ satisfies $\Gamma$}
        \STATE Commit patch $p$ and set $C_{v+1}\leftarrow C_v\oplus p$.
    \ELSE
        \STATE Store patch $p$ in Patch History $\mathcal{H}_v$.
        \STATE Keep $C_{v+1}\leftarrow C_v$.
    \ENDIF
\ENDIF
\STATE \textbf{Output:} updated controller version $C_{v+1}$ and replay-verifier report.
\end{algorithmic}
\end{algorithm}

\section{Experimental Evaluation}
\label{sec:evaluation}

\subsection{Experimental Setting and Starlink Profiles}
\label{sec:experimental-setting}
\subsubsection{Starlink Trace Processing and ABR Task}
We build the evaluation traces from the raw Starlink measurement dataset~\cite{liu2025starnet}. The raw measurements contain downlink throughput, latency, satellite candidate information, and terminal-side status logs from multiple regions. To make the traces directly replayable by ABR controllers, we convert each continuous measurement segment into a 600-s trace with 1-s throughput samples. Traces with insufficient duration, missing throughput, or invalid replay rows are discarded. Table~\ref{tab:trace-inventory} summarizes the resulting valid ABR traces. The table uses the location labels US, OSN, and VIC, corresponding to the United States, Germany, and Victoria, Canada, respectively. In the table, TP denotes throughput; P05, P50, and P95 denote the 5th-, 50th-, and 95th-percentile throughput, respectively; and the trace split is reported in the order of train/calibration/test.

\begin{table*}[!t]
\centering
\caption{Starlink ABR trace inventory.}
\label{tab:trace-inventory}
\begin{tabular}{lcccc}
\toprule
\textbf{Item} & \textbf{US} & \textbf{OSN} & \textbf{VIC} & \textbf{Total} \\
\midrule
Measurement period & 2024-04-26--2024-05-28 & 2024-07-13--2024-07-31 & 2024-07-11--2024-07-28 & 2024-04-26--2024-07-31 \\
Raw measurement runs & 16 & 118 & 188 & 322 \\
Raw throughput samples & 24,912,647 & 615,777 & 148,283 & 25,676,707 \\
Raw latency samples & 26,120,046 & 728,950 & 174,962 & 27,023,958 \\
Primary satellite changes & 23,291 & 7,982 & 1,746 & 33,019 \\
\midrule
Candidate 600-s traces & 1,955 & 1,026 & 246 & 3,227 \\
Valid 600-s traces & 1,954 & 1,024 & 243 & 3,221 \\
Filtered invalid traces & 1 & 2 & 3 & 6 \\
Trace split & 1,368/293/293 & 717/154/153 & 170/37/36 & 2,255/484/482 \\
Replay rows & 1,172,400 & 614,400 & 145,800 & 1,932,600 \\
Replay duration (h) & 325.7 & 170.7 & 40.5 & 536.8 \\
Mean TP (Mbps) & 210.1 & 216.6 & 171.7 & 209.3 \\
Mean P05 TP (Mbps) & 84.7 & 138.0 & 98.4 & 102.7 \\
Mean P50 TP (Mbps) & 223.2 & 203.1 & 171.5 & 212.9 \\
Mean P95 TP (Mbps) & 282.1 & 308.8 & 243.8 & 287.7 \\
\bottomrule
\end{tabular}
\end{table*}

All ABR experiments use the same high-bitrate video setting. The bitrate ladder is $[3,8,15,30,60,120]$ Mbps, representing a 4K/8K-style streaming service over high-capacity satellite access. Each video contains 48 chunks, and each chunk has a playback duration of 4 s~\cite{mao2017neural,huang2019comyco}. The maximum playback buffer is 60 s. The replay reward follows the standard ABR form in \eqref{eq:replay-qoe-rebuf}, with rebuffering penalty $\mu=40$ and smoothness penalty $\eta=1$.

\subsubsection{Starlink Profile Construction}
The central experimental unit is a Starlink profile rather than an isolated trace. Starting from the valid trace inventory, we construct profile batches according to region, time split, low-percentile throughput, volatility, latency, and satellite-change metadata. Table~\ref{tab:dynamic-profile-summary} summarizes the main Starlink profiles used by the replay verifier. The initial, calibration, and future-test profiles model profile arrival over time. The high-volatility, low-tail, and high-latency profiles expose different failure patterns. The regression memory profile is selected from comparatively stable traces and is used only to check whether a patch damages previously safe operating conditions.

\begin{table*}[!t]
\centering
\caption{Summary of Starlink profiles used for controller maintenance.}
\label{tab:dynamic-profile-summary}
\begin{tabular}{lcccccc}
\toprule
\textbf{Profile} & \textbf{Traces} & \textbf{Region Mix (US/OSN/VIC)} & \textbf{Mean TP} & \textbf{Mean P05 TP} & \textbf{Volatility} & \textbf{Latency} \\
 & & & \textbf{(Mbps)} & \textbf{(Mbps)} & \textbf{(Mbps)} & \textbf{(ms)} \\
\midrule
Initial deployment & 192 & 64/64/64 & 190.9 & 91.7 & 195.1 & 38.8 \\
New calibration & 484 & 293/154/37 & 200.4 & 88.7 & 205.3 & 29.7 \\
High volatility & 192 & 173/18/1 & 197.4 & 54.7 & 251.1 & 30.2 \\
Low tail & 192 & 126/50/16 & 123.1 & 18.9 & 211.3 & 38.8 \\
High latency & 192 & 106/47/39 & 123.1 & 28.2 & 199.1 & 41.0 \\
Future test & 482 & 293/153/36 & 192.7 & 85.9 & 198.4 & 30.7 \\
Regression memory & 192 & 64/64/64 & 277.3 & 195.9 & 155.9 & 28.1 \\
\bottomrule
\end{tabular}
\end{table*}

\subsubsection{Controllers, Variants, and Implementation}
We evaluate SafeLink-Agent on both rule-based and learned ABR controllers. The rule-based branch includes BOLA and RobustMPC. BOLA is controlled by buffer thresholds, and RobustMPC is controlled by the prediction horizon and safety margin. The Static default variant uses one default configuration for all profiles. The Static conservative variant applies one safer configuration globally. SafeLink-Agent instead selects a verified profile-conditioned patch, so that safer configurations are used only on profiles where replay evidence indicates a severe-risk reduction without unacceptable regression.

For BOLA, the default configuration uses a 10-s minimum buffer and a 30-s target buffer, while the conservative configuration uses a 15-s minimum buffer and a 45-s target buffer. For RobustMPC, the default configuration uses a five-chunk planning horizon with safety margin 1.00. The main conservative configuration uses a six-chunk horizon with safety margin 0.85; additional horizon and margin candidates are used only when constructing verifier upper bound or ablation variants.

The learned controller branch includes SABR, Pensieve, and Comyco. For these controllers, SafeLink-Agent attaches a policy-agnostic runtime auditor after the learned policy. We compare three variants: No auditor, Static auditor, and Adaptive auditor. 

In our implementation, SafeLink-Agent uses DeepSeek-V4-Pro~\cite{deepseek2026deepseek} as the LLM in the agentic patch proposer. The verifier constraints $\Gamma$ are instantiated with the severe-session threshold $b_{\mathrm{sev}}=10$ s and tail fraction $\alpha=0.05$. For the maintenance trigger in \eqref{eq:maintenance-trigger}, we set $\beta_{\mathrm{sev}}=1\%$ and $\beta_{\mathrm{tail}}=10$ s.

For rule-based controller patches, target-profile replay instantiates the target block $\Gamma_{\mathrm{tar}}$ in \eqref{eq:target-constraints}: the selected configuration must keep $R_{\mathrm{sev}}(C_v\oplus p,D_k)\leq 1\%$, reduce at least one target-profile risk metric, i.e., $R_{\mathrm{sev}}$ or $T_{\alpha}$, and keep the target-profile QoE loss within $\epsilon=8\%$. Regression-memory replay instantiates $\Gamma_{\mathrm{mem}}$ in \eqref{eq:memory-constraints} with $\epsilon_{\mathrm{reg}}=2\%$ and $\zeta=0.5$ percentage points; it also applies a worst-5\% rebuffering-increase tolerance of 1 s.

For learned controller auditor patches, target-profile replay uses $\delta=0.5$ percentage points for severe-session reduction, requires non-increasing $T_{\alpha}$, and checks that more severe sessions are fixed than newly introduced. The QoE-cost bound $\epsilon$ is set to 3\% for SABR and 5\% for Pensieve and Comyco; the corresponding audit-rate increase limits are 15\% and 35\%, respectively. Regression-memory replay uses the same QoE and audit limits, requires no severe-session increase, i.e., $\zeta=0$, and allows only 0.1 s tolerance in worst-5\% rebuffering.

\begin{table}[!t]
\centering
\caption{Main controller maintenance variants.}
\label{tab:maintenance-variants}
\begin{tabular}{lp{0.58\linewidth}}
\toprule
\textbf{Variant} & \textbf{Description} \\
\midrule
Static default & Original controller configuration or policy used for all profiles. \\
Static conservative & One globally safer configuration used for all profiles. \\
SafeLink-Agent & Profile-conditioned patch accepted by target-profile replay and regression-memory replay. \\
Verifier upper bound & Best feasible profile-wise configuration selected with replay evidence; used only for analysis. \\
Static auditor & Fixed runtime safety layer attached to a learned ABR policy. \\
Adaptive auditor & Runtime safety layer whose guard changes with profile or failure context. \\
\bottomrule
\end{tabular}
\end{table}

\subsubsection{Evaluation Metrics}

For a profile $D$ with $|D|$ replay sessions, we report the average QoE and mean cumulative rebuffering:
\begin{equation}
    \bar{Q}(C,D)=\frac{1}{|D|}\sum_{\tau\in D}Q(C,\tau),
    \quad
    \bar{B}(C,D)=\frac{1}{|D|}\sum_{\tau\in D}B(C,\tau).
    \label{eq:eval-average-metrics}
\end{equation}
To quantify severe session-level risk, we report the fraction of sessions whose cumulative rebuffering exceeds 10 s:
\begin{equation}
    S_{>10}(C,D)
    =
    \frac{1}{|D|}
    \sum_{\tau\in D}
    \mathbf{1}\{B(C,\tau)>10\}.
    \label{eq:eval-severe-ratio}
\end{equation}
We also report worst-5\% session rebuffering, denoted by $B_{\mathrm{worst5}}$, which is the average cumulative rebuffering of the 5\% sessions with the largest $B(C,\tau)$. For maintenance analysis, we additionally report QoE loss relative to the Static default controller, QoE loss on regression memory, audit rate for learned controller auditors, and verifier acceptance or rejection.

\subsection{Rule-Based Controller Maintenance}
\label{sec:rule-based-maintenance}
This experiment evaluates whether SafeLink-Agent can maintain rule-based ABR controllers across newly observed Starlink profiles. The comparison is designed around the deployment choices faced by a fixed controller. The Static default controller preserves its original QoE-oriented configuration on every profile. The Static conservative controller applies one safer configuration globally. SafeLink-Agent applies a profile-conditioned patch only when replay verification indicates that the target-profile risk reduction does not introduce regression on profiles in regression memory.

\begin{figure}[!t]
  \centering
  \includegraphics[width=\linewidth]{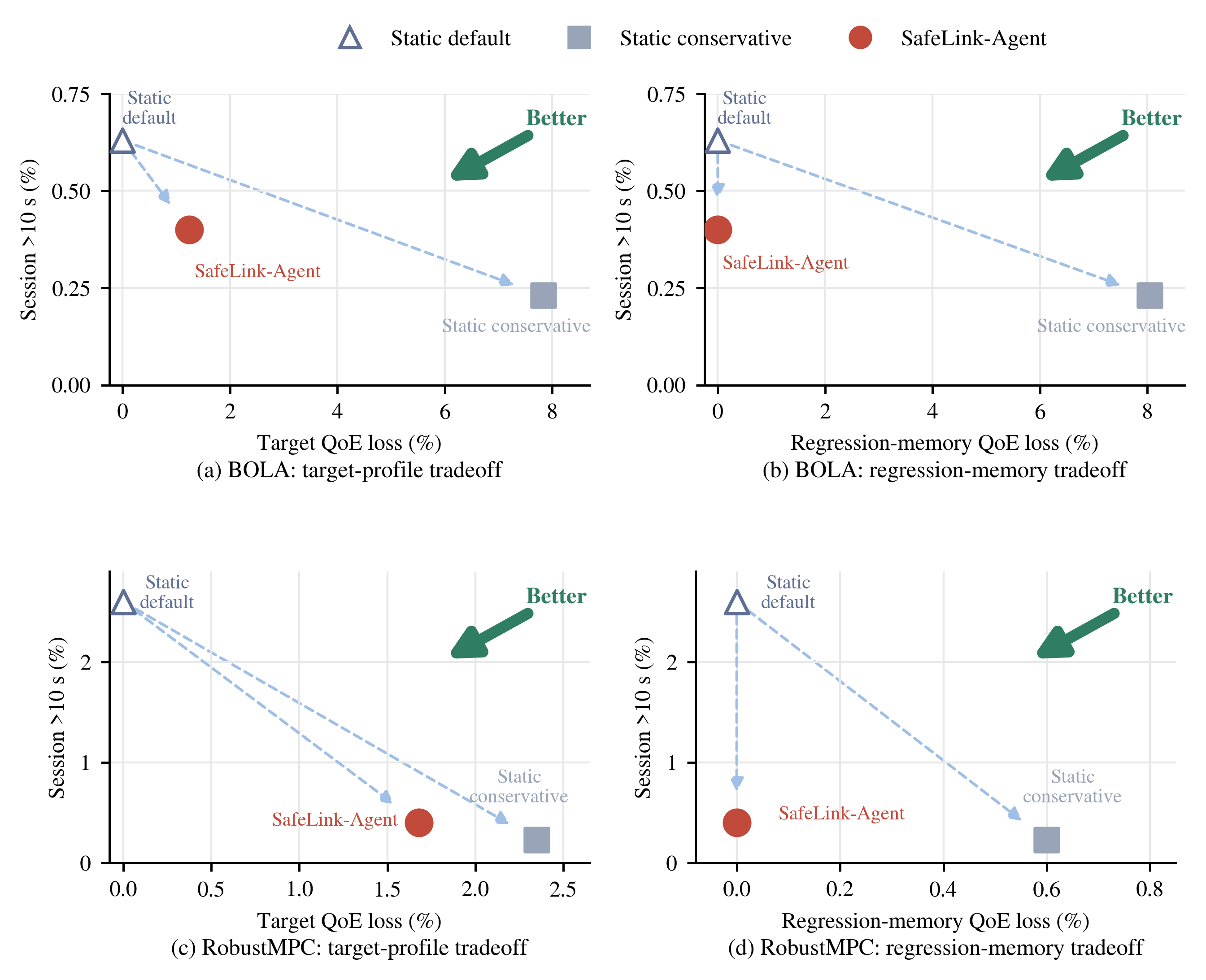}
  \caption{Rule-based controller maintenance tradeoff.}
  \label{fig:rule-based-tradeoff}
\end{figure}

Fig.~\ref{fig:rule-based-tradeoff} visualizes the rule-based result as a QoE loss versus severe-risk tradeoff. In each panel, the lower-left direction is preferred: a good maintenance outcome should reduce severe-session risk while keeping target-profile QoE loss or QoE loss on regression memory small. Table~\ref{tab:rule-based-main} reports the corresponding exact values over all non-regression Starlink profiles, including initial, calibration, high-volatility, low-tail, high-latency, and future-test profiles.

\begin{table*}[!t]
\centering
\caption{Rule-based controller maintenance across Starlink profiles.}
\label{tab:rule-based-main}
\begin{tabular}{llcccccc}
\toprule
\textbf{Controller} & \textbf{Method} & \textbf{QoE} & \textbf{QoE Loss} & \textbf{Mean Rebuf} & \textbf{Worst-5\% Rebuf} & \textbf{Session $>10$s} & \textbf{Reg. QoE Loss} \\
 & & & \textbf{(\%)} & \textbf{(s)} & \textbf{(s)} & \textbf{(\%)} & \textbf{(\%)} \\
\midrule
BOLA & Static default & 87.42 & 0.00 & 0.70 & 7.23 & 0.63 & 0.00 \\
BOLA & Static conservative & 80.57 & 7.84 & 0.55 & 4.26 & 0.23 & 8.05 \\
BOLA & \textbf{SafeLink-Agent} & \textbf{86.34} & \textbf{1.24} & \textbf{0.62} & \textbf{5.63} & \textbf{0.40} & \textbf{0.00} \\
\midrule
RobustMPC & Static default & 96.58 & 0.00 & 1.08 & 11.77 & 2.60 & 0.00 \\
RobustMPC & Static conservative & 94.31 & 2.35 & 0.61 & 4.97 & 0.23 & 0.60 \\
RobustMPC & \textbf{SafeLink-Agent} & \textbf{94.95} & \textbf{1.68} & \textbf{0.66} & \textbf{5.70} & \textbf{0.40} & \textbf{0.00} \\
\bottomrule
\end{tabular}
\end{table*}

Fig.~\ref{fig:rule-based-tradeoff} shows that Static conservative tuning and SafeLink-Agent move in the same risk reduction direction from the Static default point, but with different side effects. In the BOLA panels, the Static conservative point is far to the right, indicating a large QoE cost for a relatively small additional reduction in severe-session risk. SafeLink-Agent stays closer to the low-loss side of the plot while still moving below the default severe-risk level. The exact values in Table~\ref{tab:rule-based-main} show that Static conservative tuning reduces the BOLA severe-session ratio from 0.63\% to 0.23\%, but causes 7.84\% target-profile QoE loss and 8.05\% QoE loss on regression memory. SafeLink-Agent reaches a 0.40\% severe-session ratio with only 1.24\% target-profile QoE loss and zero QoE loss on regression memory.

The RobustMPC panels make the same tradeoff more visible because the Static default controller has a clearer severe-session tail. In Fig.~\ref{fig:rule-based-tradeoff}, SafeLink-Agent moves the operating point close to the low-risk side reached by Static conservative tuning, but with smaller target-profile QoE loss and no QoE loss on regression memory. Numerically, SafeLink-Agent reduces the RobustMPC severe-session ratio from 2.60\% to 0.40\%, an 84.6\% reduction over the Static default. Compared with Static conservative tuning, it obtains 92.8\% of the severe-risk reduction while lowering the target QoE loss from 2.35\% to 1.68\% and eliminating QoE loss on regression memory from 0.60\% to 0.00\%. These results show that SafeLink-Agent is not a Static conservative fallback; it uses profile-conditioned patches accepted by replay verification to reduce severe risk while controlling QoE loss and avoiding regression on memory profiles.

\subsection{Learned Controller Maintenance}
\label{sec:learned-maintenance}
This experiment evaluates whether SafeLink-Agent can maintain learned ABR controllers without retraining them from scratch. SafeLink-Agent uses a policy-agnostic runtime auditor as the patch type: the learned policy still proposes the bitrate, and the auditor modifies only requests that violate the verified safety rule.

\begin{figure}[!t]
  \centering
  \includegraphics[width=\linewidth]{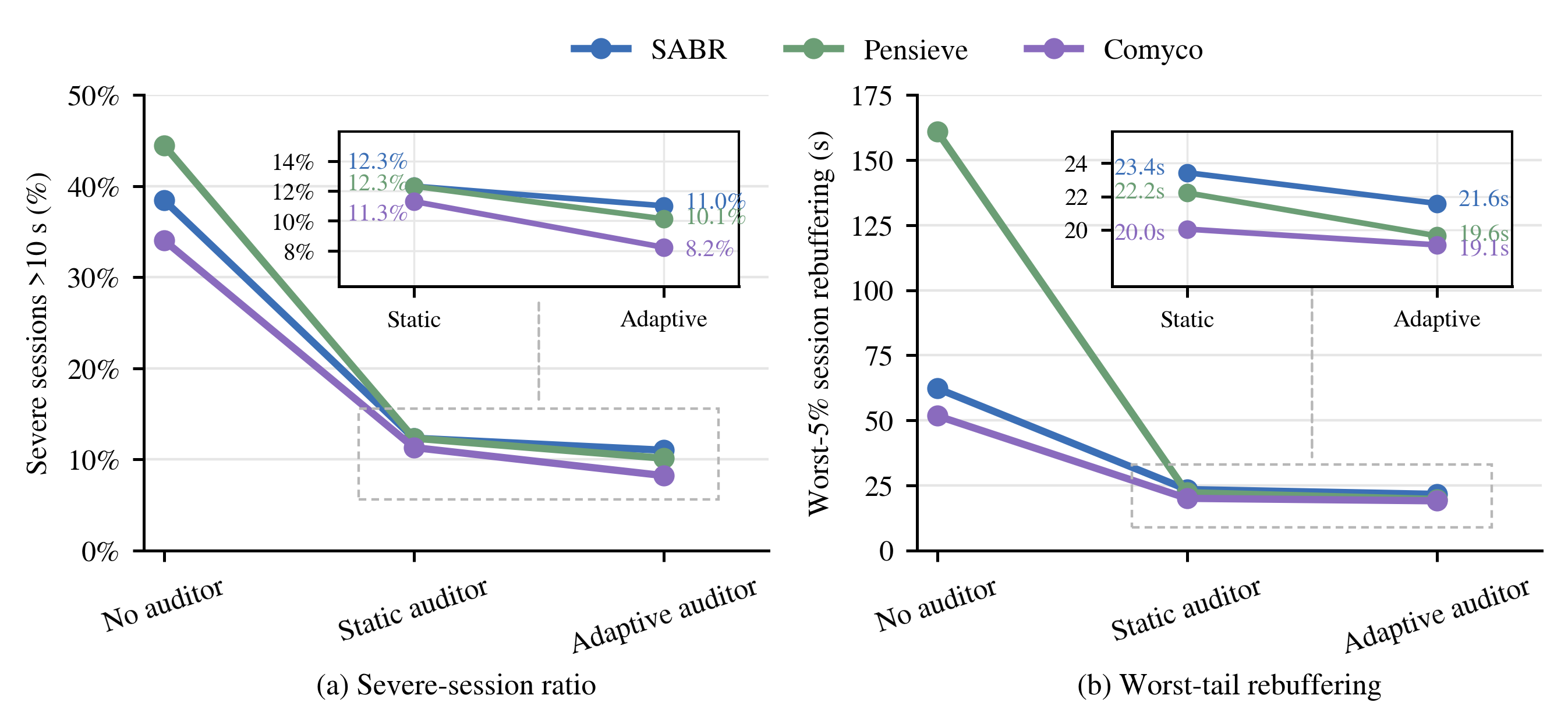}
  \caption{Learned controller tail-risk reduction with verified runtime auditors.}
  \label{fig:learned-tail-reduction}
\end{figure}

We compare SABR, Pensieve, and Comyco under three operating modes. The No auditor mode evaluates the original learned controller. The Static auditor mode attaches one verified safety rule to all evaluated traces. The Adaptive auditor mode further conditions the guard on failure context, such as low buffer or high-risk link states. Fig.~\ref{fig:learned-tail-reduction} visualizes the severe-session and worst-tail rebuffering trends, while Table~\ref{tab:learned-main} reports the exact values under the default severe-stall threshold of 10 s and the worst-5\% tail metric.

\begin{table*}[!t]
\centering
\caption{Learned controller maintenance with verified runtime auditors.}
\label{tab:learned-main}
\begin{tabular}{llccccc}
\toprule
\textbf{Controller} & \textbf{Method} & \textbf{QoE} & \textbf{Mean Rebuf} & \textbf{Worst-5\% Rebuf} & \textbf{Session $>10$s} & \textbf{Audit Rate} \\
 & & & \textbf{(s)} & \textbf{(s)} & \textbf{(\%)} & \textbf{(\%)} \\
\midrule
SABR & No auditor & 4372.95 & 12.91 & 62.31 & 38.47 & 0.00 \\
SABR & Static auditor & 4311.30 & 3.19 & 23.43 & 12.33 & 11.26 \\
SABR & \textbf{Adaptive auditor} & \textbf{4271.32} & \textbf{2.65} & \textbf{21.58} & \textbf{11.01} & \textbf{12.48} \\
\midrule
Pensieve & No auditor & 4129.85 & 30.39 & 161.05 & 44.49 & 0.00 \\
Pensieve & Static auditor & 4290.27 & 3.26 & 22.22 & 12.33 & 21.39 \\
Pensieve & \textbf{Adaptive auditor} & \textbf{4234.21} & \textbf{2.52} & \textbf{19.64} & \textbf{10.13} & \textbf{21.93} \\
\midrule
Comyco & No auditor & 4427.39 & 10.08 & 51.77 & 34.07 & 0.00 \\
Comyco & Static auditor & 4339.09 & 2.81 & 20.04 & 11.31 & 8.11 \\
Comyco & \textbf{Adaptive auditor} & \textbf{4290.84} & \textbf{2.33} & \textbf{19.09} & \textbf{8.22} & \textbf{9.55} \\
\bottomrule
\end{tabular}
\end{table*}

In Fig.~\ref{fig:learned-tail-reduction}, the No auditor learned controllers expose a pronounced severe-session tail on hard Starlink profiles. SABR, Pensieve, and Comyco have severe-session ratios of 38.47\%, 44.49\%, and 34.07\%, respectively. Their worst-5\% session rebuffering is also large, especially for Pensieve, whose worst-5\% rebuffering reaches 161.05 s. After SafeLink-Agent attaches a verified Static auditor, the severe-session ratio decreases to 12.33\%, 12.33\%, and 11.31\% for SABR, Pensieve, and Comyco, respectively. The Adaptive auditor further reduces the corresponding ratios to 11.01\%, 10.13\%, and 8.22\%.

Adaptive auditing usually gives the lowest severe-session risk and mean rebuffering, but it increases the audit rate and can reduce QoE compared with the Static auditor. For example, on SABR, the Adaptive auditor reduces severe sessions from 12.33\% to 11.01\% compared with the Static auditor, but QoE decreases from 4311.30 to 4271.32, and the audit rate increases from 11.26\% to 12.48\%. On Pensieve, the auditor improves both QoE and severe risk relative to the No auditor baseline, because avoiding long stalls outweighs the bitrate corrections. These differences support the SafeLink-Agent design: the proposer generates a candidate auditor patch, while replay verification determines whether the resulting QoE--risk operating point is acceptable for deployment.

\subsection{Rolling Maintenance Across Newly Observed Starlink Profiles}
\label{sec:rolling-maintenance}
This experiment evaluates the maintenance process when Starlink profiles arrive sequentially. The evaluated sequence is initial deployment, new calibration traces, high-volatility traces, low-tail traces, high-latency traces, and future-test traces. At each step, the Static default controller keeps its original configuration, the Static conservative controller keeps one global safe configuration, and SafeLink-Agent uses only verifier-accepted profile-conditioned patches.

\begin{figure}[!t]
  \centering
  \includegraphics[width=\linewidth]{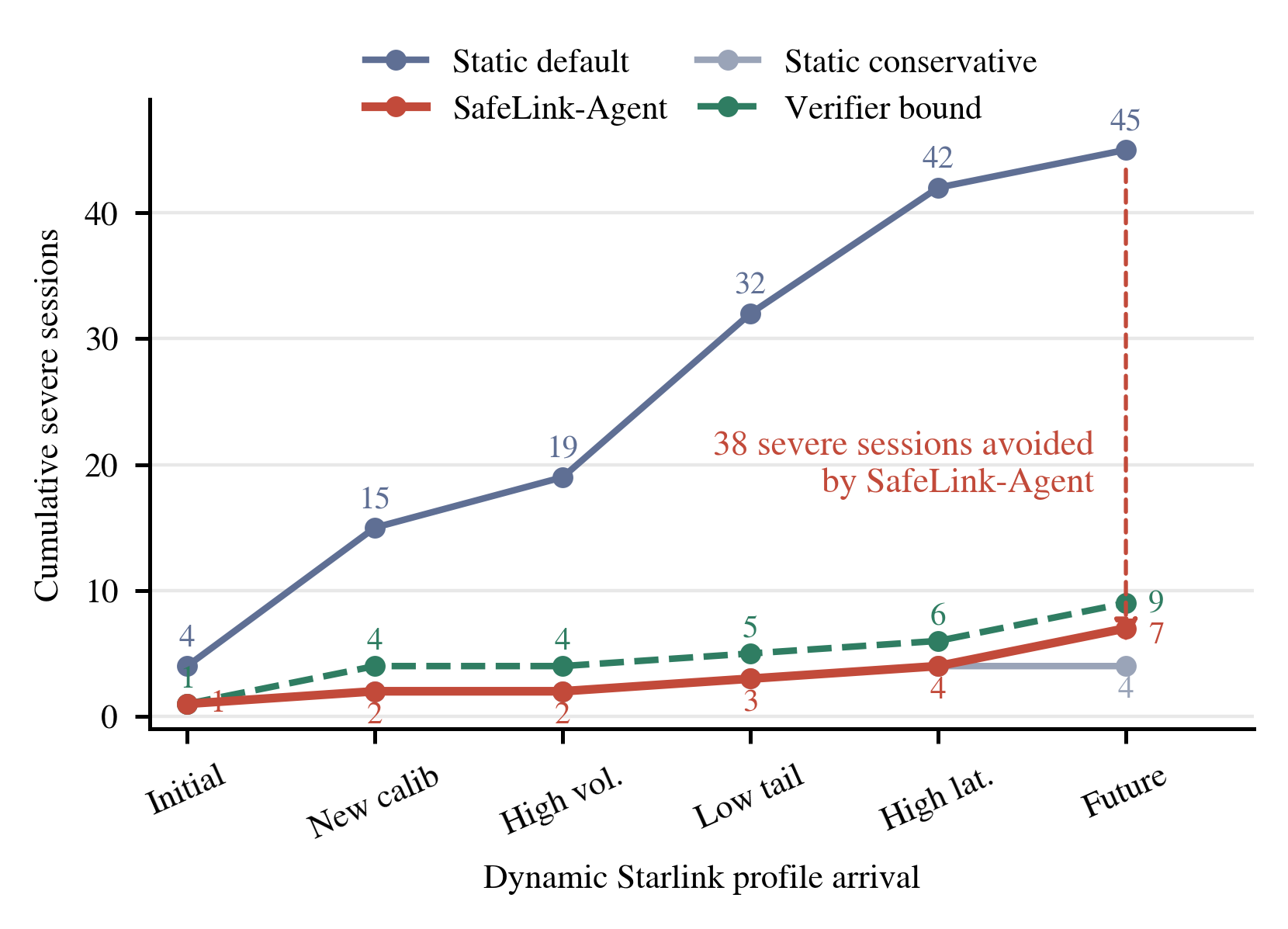}
  \caption{Rolling maintenance across newly observed Starlink profiles.}
  \label{fig:rolling-maintenance}
\end{figure}

Fig.~\ref{fig:rolling-maintenance} plots the cumulative number of severe sessions as profiles arrive. The terminal labels report the final cumulative severe-session counts, and the verifier upper bound is included only as an analysis reference; it selects the best risk-feasible configuration from the existing replay sweep and is not treated as a deployable controller.

The Static default controller accumulates severe sessions quickly once low-tail and high-latency profiles arrive. Its cumulative severe-session count increases from 19 after the high-volatility profile to 32 after the low-tail profile and to 42 after the high-latency profile, ending with 45 severe sessions. This trend shows why a one-time controller configuration can become unsafe under later Starlink profiles even if it preserves high average QoE.

The Static conservative controller ends with only four severe sessions, but it reaches this point by applying one safe configuration to every profile. SafeLink-Agent ends with seven severe sessions, reducing 38 severe sessions relative to the Static default controller while remaining close to the verifier upper bound of nine severe sessions. The rolling trend supports the central maintenance view of this paper: the controller should not be globally frozen or statically conservative, but should be patched only when new profile evidence justifies the change.

\subsection{Handover-Heavy Starlink Stress Test}
\label{sec:handover-heavy}
This experiment evaluates whether the maintenance need is connected to Starlink-specific mobility. We use candidate-satellite change metadata as a serving-satellite-change proxy. Among all non-regression dynamic sessions, the top 30\% by primary candidate-satellite changes are labeled handover-heavy, and the remaining sessions are labeled normal. This split produces 521 handover-heavy sessions and 1213 normal sessions. The handover-heavy group has many more primary candidate-satellite changes on average, 1924.71 versus 758.52, and also has a lower mean P05 throughput, 63.83 Mbps versus 72.77 Mbps. Thus, the stress set captures sessions with more frequent satellite-association changes and a weaker throughput tail.

\begin{figure}[!t]
  \centering
  \includegraphics[width=\linewidth]{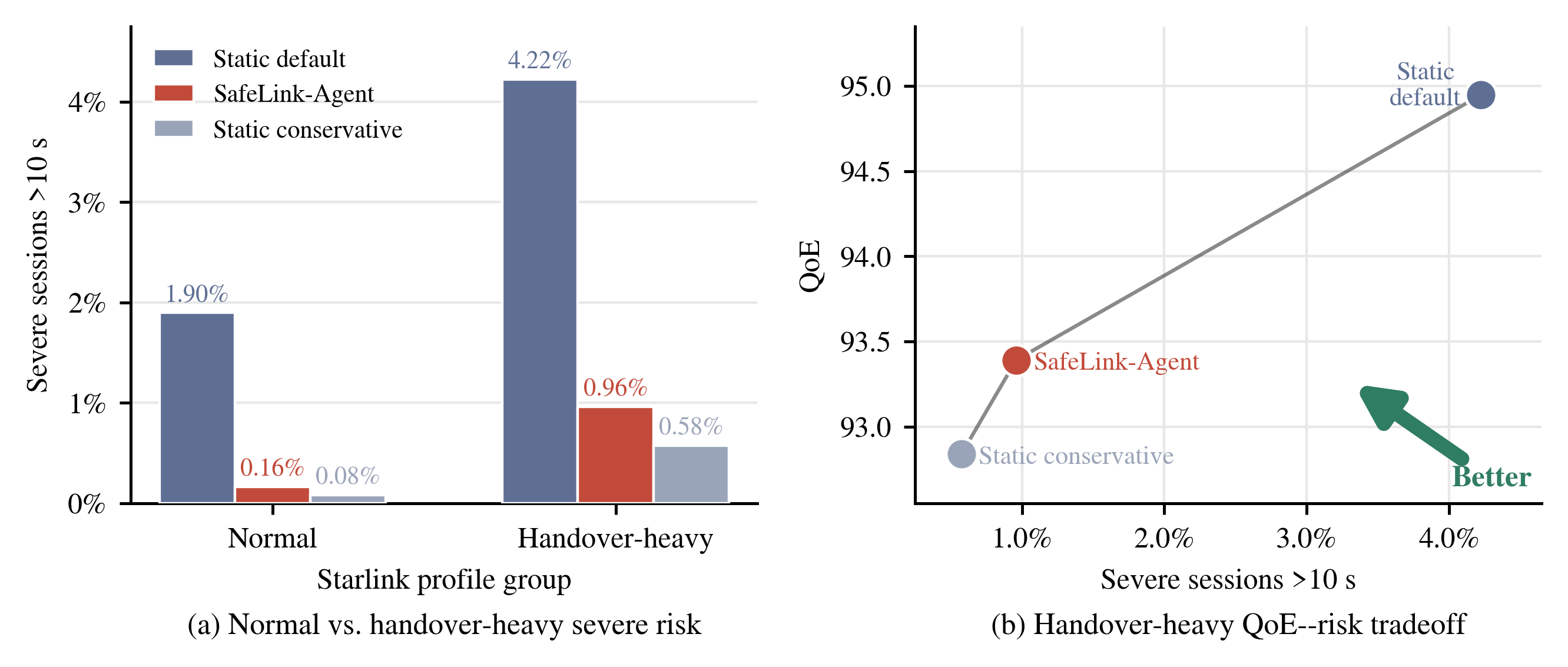}
  \caption{Stress test on handover-heavy Starlink profiles.}
  \label{fig:handover-heavy}
\end{figure}

Fig.~\ref{fig:handover-heavy} summarizes the RobustMPC result. Panel (a) compares the severe-session ratio on normal and handover-heavy sessions, while panel (b) shows the handover-heavy QoE--risk operating points. The Static default controller is sensitive to the mobility-heavy split: its severe-session ratio increases from 1.90\% on normal sessions to 4.22\% on handover-heavy sessions. This confirms that handover-heavy Starlink periods are not merely another random subset; they amplify the severe-session tail of a high-QoE rule-based controller.

SafeLink-Agent reduces the handover-heavy severe-session ratio from 4.22\% to 0.96\%. The Static conservative baseline gives the lowest severe-session ratio, 0.58\%, but panel (b) shows that it occupies a lower-QoE operating point than SafeLink-Agent on the handover-heavy split, 92.84 versus 93.39. In the rule-based main result, the Static conservative configuration introduces QoE loss on regression memory, whereas SafeLink-Agent keeps this loss at zero. The handover-heavy result therefore supports the maintenance: when mobility changes expose a hard profile, a verified patch can reduce severe stalls without turning the entire controller into a Static conservative version.

\subsection{Patch Verification and Regression Memory Analysis}
\label{sec:internal-analysis}
This experiment examines the acceptance logic behind SafeLink-Agent. Since candidate patches may look reasonable on a small number of traces, the key question is whether SafeLink-Agent can reject proposals that fail under full replay or create regressions on previously stable profiles. We therefore analyze two mechanisms: replay verification, which tests whether a candidate patch satisfies the target-profile constraints, and regression memory, which prevents a patch from improving the current profile at the cost of previously validated behavior.

\subsubsection{Verifier Ablation}
This ablation tests whether a small probe check can replace full replay verification. The probe check runs a candidate RobustMPC patch on a small trace subset, while the full verifier replays the same patch on the complete target profile. As shown in Table~\ref{tab:verifier-ablation}, the candidate passes the probe on all four profiles. Full replay changes the decision: the patch is rejected on the high-latency and low-tail profiles because their severe-session ratios remain above the verifier constraint. Therefore, a probe can screen proposals, but it cannot decide patch commitment.

\begin{table}[!t]
\centering
\caption{Verifier ablation for a probe-accepted RobustMPC proposal.}
\label{tab:verifier-ablation}
\footnotesize
\begin{tabular}{@{}lcccc@{}}
\toprule
\textbf{Metric} & \textbf{High Lat.} & \textbf{High Vol.} & \textbf{Low Tail} & \textbf{Initial} \\
\midrule
Probe pass & Yes & Yes & Yes & Yes \\
Full pass & No & Yes & No & Yes \\
Session $>10$s (\%) & 2.60 & 0.00 & 2.60 & 0.52 \\
QoE delta (\%) & -1.37 & -1.73 & -1.45 & -1.80 \\
Reject reason & Severe & -- & Severe & -- \\
\bottomrule
\end{tabular}
\end{table}

\subsubsection{Regression Memory Ablation}
The regression memory analysis evaluates whether a target-profile repair creates unacceptable side effects on previously stable profiles. Table~\ref{tab:regression-memory} compares the default controller, a target-only safe patch, and the SafeLink-Agent patch accepted with regression memory. For BOLA, the target-only patch reduces the target severe-session ratio from 0.63\% to 0.23\%, but it introduces an 8.05\% QoE regression on the memory set. SafeLink-Agent accepts a less aggressive memory-aware patch: it still reduces target risk relative to the default controller, while keeping QoE loss on regression memory at zero. RobustMPC shows the same pattern with a smaller loss on regression memory. These results indicate that regression memory changes the accepted patch, rather than only adding an after-the-fact report.

\begin{table*}[!t]
\centering
\caption{Regression memory analysis for target-profile repair.}
\label{tab:regression-memory}
\begin{tabular}{llcccc}
\toprule
\textbf{Controller} & \textbf{Variant} & \textbf{Target QoE} & \textbf{Worst-5\% Rebuf} & \textbf{Session $>10$s} & \textbf{Reg. QoE Loss} \\
 & & & \textbf{(s)} & \textbf{(\%)} & \textbf{(\%)} \\
\midrule
BOLA & Static default & 87.42 & 7.23 & 0.63 & 0.00 \\
BOLA & Target-only safe patch & 80.57 & 4.26 & 0.23 & 8.05 \\
BOLA & SafeLink-Agent with memory & 86.34 & 5.63 & 0.40 & 0.00 \\
\midrule
RobustMPC & Static default & 96.58 & 11.77 & 2.60 & 0.00 \\
RobustMPC & Target-only safe patch & 94.31 & 4.97 & 0.23 & 0.60 \\
RobustMPC & SafeLink-Agent with memory & 94.95 & 5.70 & 0.40 & 0.00 \\
\bottomrule
\end{tabular}
\end{table*}

Together, these results show that both checks affect the final maintenance decision. Full replay rejects patches that pass a small probe but still violate severe-risk constraints, while regression memory changes an over-conservative target-only patch into a lower-regression patch. This ablation confirms that removing either full replay verification or regression-memory replay can lead to unsafe or unnecessarily regressive patch acceptance.

\subsection{Hard Trace Mechanism Case Study}
\label{sec:case-study}
\begin{figure}[!t]
  \centering
  \includegraphics[width=\linewidth]{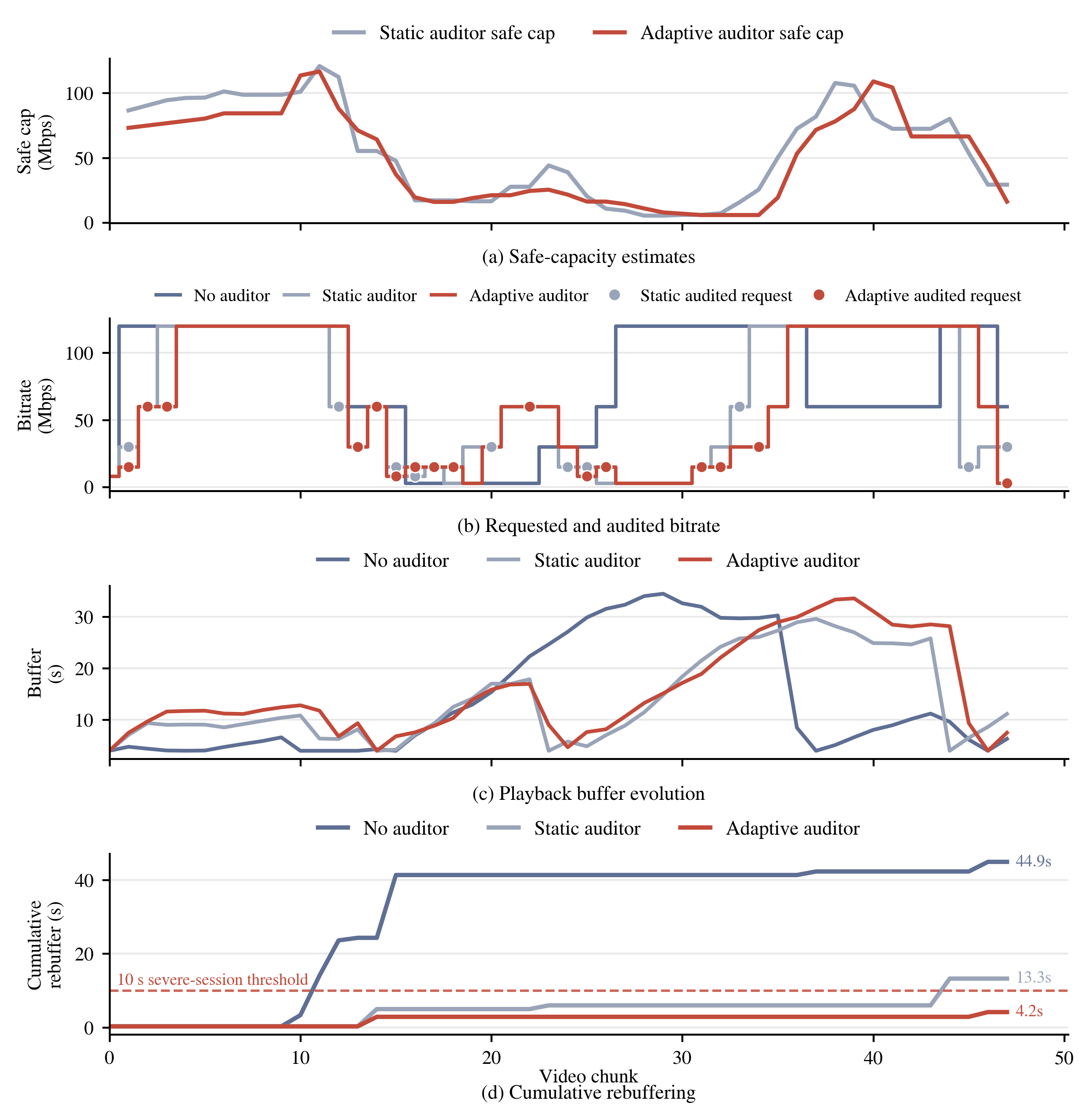}
  \caption{Mechanism case study on a hard learned controller session.}
  \label{fig:case-study-timeline}
\end{figure}

This experiment inspects one hard learned controller session to explain the mechanism behind the aggregate severe-risk reduction. The selected session comes from a low-tail OSN profile where the original learned policy produces repeated large rebuffering events. We compare the same replay trace under three variants: the original learned controller without auditing, a verified Static auditor, and the Adaptive auditor accepted by SafeLink-Agent. Fig.~\ref{fig:case-study-timeline} visualizes the chunk-level behavior, and Table~\ref{tab:case-study-summary} reports the corresponding session-level outcome.

\begin{table}[!t]
\centering
\caption{Hard trace case study for a learned controller auditor patch.}
\label{tab:case-study-summary}
\footnotesize
\begin{tabular}{@{}lccccc@{}}
\toprule
\textbf{Variant} & \textbf{QoE} & \textbf{Rebuf} & \textbf{Max Chunk} & \textbf{Bitrate} & \textbf{Audit} \\
 & & \textbf{(s)} & \textbf{(s)} & \textbf{(Mbps)} & \textbf{(\%)} \\
\midrule
No auditor & 1518.96 & 44.85 & 17.00 & 79.98 & 0.00 \\
Static auditor & 1969.76 & 13.26 & 7.25 & 65.71 & 29.17 \\
Adaptive auditor & 2224.30 & 4.19 & 2.59 & 63.69 & 33.33 \\
\bottomrule
\end{tabular}
\end{table}

Fig.~\ref{fig:case-study-timeline} shows how the accepted auditor patch changes the session trajectory. Panel (a) first shows the safe-capacity estimates used by the Static and Adaptive auditors. Both estimates expose a long low-capacity interval in the middle of the session and a second capacity drop near the end, which creates the conditions for severe rebuffering if high bitrates are admitted. Panel (b) then shows the requested and audited bitrates. The No auditor rollout keeps several high-bitrate requests during the hard intervals, whereas the Static and Adaptive auditors replace part of these requests with lower audited bitrates.

The playback consequence appears in panels (c) and (d). Panel (c) shows that the No auditor rollout stays close to the startup-buffer level during the early hard interval, which corresponds to the sharp cumulative-rebuffering increase in panel (d). The audited rollouts build more buffer before this early stall period, so their cumulative rebuffering remains much lower at the same chunks. The Static auditor, however, later suffers a pronounced buffer drop near the end of the session; the Adaptive auditor also loses buffer near the end, but the added stall is smaller. Panel (d) summarizes this difference. The No auditor rollout crosses the 10-s severe-session threshold early and reaches 44.85 s cumulative rebuffering. The Static auditor delays and reduces the stall, but still ends above the threshold at 13.26 s. The Adaptive auditor remains below the threshold and ends at 4.19 s, turning the same hard trace into a non-severe session.

Table~\ref{tab:case-study-summary} reports the corresponding session-level outcome. Compared with the Static auditor, the Adaptive auditor increases the intervention rate from 29.17\% to 33.33\%, reduces total rebuffering from 13.26 s to 4.19 s, reduces the maximum single-chunk rebuffering from 7.25 s to 2.59 s, and improves QoE from 1969.76 to 2224.30. These values support the visual observation that a small increase in verified auditing can materially change the stall outcome on a hard learned controller session.

The case study shows that the learned controller patch acts as a targeted runtime auditor. It preserves the original policy as the bitrate proposer, but uses the maintenance context to identify requests that become risky under low buffer or low safe capacity. The behavior explains the aggregate learned controller result: verified auditor patches can reduce severe-session tails by correcting a small set of high-risk decisions rather than replacing the underlying ABR policy.

\subsection{Sensitivity Analysis}
\label{sec:sensitivity-analysis}
This experiment evaluates whether the maintenance conclusion depends on a single severe-stall threshold or tail fraction. We vary the severe-session threshold from 5 s to 15 s and the worst-tail fraction from 1\% to 10\%. Table~\ref{tab:sensitivity-summary} reports a compact summary. For rule-based controllers, we use RobustMPC because it is the clearest rule-based case in the main result. For learned controllers, we report the average over SABR, Pensieve, and Comyco to summarize the effect of the verified Adaptive auditor.

\begin{table*}[!t]
\centering
\caption{Sensitivity to severe-stall threshold and worst-tail fraction.}
\label{tab:sensitivity-summary}
\begin{tabular}{llcccccc}
\toprule
\textbf{Branch} & \textbf{Method} & \textbf{$>5$s} & \textbf{$>10$s} & \textbf{$>15$s} & \textbf{Worst-1\%} & \textbf{Worst-5\%} & \textbf{Worst-10\%} \\
 & & \textbf{(\%)} & \textbf{(\%)} & \textbf{(\%)} & \textbf{(s)} & \textbf{(s)} & \textbf{(s)} \\
\midrule
RobustMPC & Static default & 4.73 & 2.60 & 1.04 & 22.17 & 11.77 & 7.32 \\
RobustMPC & SafeLink-Agent & 1.90 & 0.40 & 0.29 & 12.69 & 5.70 & 3.59 \\
\midrule
Learned avg. & No auditor & 48.60 & 39.01 & 32.65 & 117.52 & 91.71 & 78.29 \\
Learned avg. & Adaptive auditor & 13.80 & 9.79 & 4.16 & 26.71 & 20.10 & 16.03 \\
\bottomrule
\end{tabular}
\end{table*}

The absolute severe-session ratio changes with the threshold, as expected, but the ordering remains stable. For RobustMPC, SafeLink-Agent reduces the severe-session ratio from 4.73\% to 1.90\% at the 5-s threshold, from 2.60\% to 0.40\% at the 10-s threshold, and from 1.04\% to 0.29\% at the 15-s threshold. The worst-tail metric shows the same trend under different tail fractions. For the learned controllers, the Adaptive auditor reduces the average severe-session ratio from 39.01\% to 9.79\% at the default 10-s threshold, and the reduction remains visible under both stricter and looser thresholds. These results show that the maintenance gain is robust to reasonable choices of severe-stall threshold and tail-risk fraction.

\section{Conclusion}
This paper presented SafeLink-Agent, an agentic maintenance framework for adaptive bitrate controllers over dynamic Starlink networks. We formulated Starlink ABR as a controller-maintenance problem, where newly observed profiles expose failures that should be repaired through controller updates under severe-risk, QoE cost, and regression constraints. SafeLink-Agent realizes this formulation by mining replay evidence, using an agentic patch proposer to generate candidate patches, and committing only the patches that pass target-profile replay and regression-memory replay. This turns newly observed Starlink failures into verifiable maintenance actions that keep ABR controllers robust as the access environment changes. Experiments on real Starlink traces show that SafeLink-Agent reduces the RobustMPC severe-session ratio from 2.60\% to 0.40\%, lowers cumulative severe sessions from 45 to 7 in rolling maintenance, and reduces the learned controller average severe-session ratio from 39.01\% to 9.79\%. These results show that SafeLink-Agent can combine LLM-assisted patch generation with replay verification for automated maintenance and localized repair of ABR controllers, reducing severe-session failures without repeated manual retuning or retraining under dynamic Starlink access. Future work will investigate broader agentic maintenance across prediction, control, and application-layer optimization modules in satellite networking systems.

\bibliographystyle{IEEEtran}
\bibliography{ref}

\clearpage
\section*{Supplementary Material}
\setcounter{section}{0}
\renewcommand{\thesection}{S\arabic{section}}
\setcounter{table}{0}
\renewcommand{\thetable}{S\arabic{table}}
\setcounter{figure}{0}
\renewcommand{\thefigure}{S\arabic{figure}}
\setcounter{equation}{0}
\renewcommand{\theequation}{S\arabic{equation}}

\section{Main Notation}

Table~\ref{tab:notation} summarizes the notation used in the controller-maintenance formulation and SafeLink-Agent design.

\begin{table*}[!t]
\centering
\caption{Main notation used in controller maintenance.}
\label{tab:notation}
\footnotesize
\setlength{\tabcolsep}{3pt}
\begin{tabular}{@{}p{0.18\textwidth}p{0.30\textwidth}p{0.18\textwidth}p{0.30\textwidth}@{}}
\toprule
\textbf{Symbol} & \textbf{Meaning} & \textbf{Symbol} & \textbf{Meaning} \\
\midrule
$t$, $T$ & Chunk index and number of chunks in one streaming session. & $\tau_{k,j}$, $D_k$, $n_k$ & The $j$th replay session, the $k$th Starlink profile, and the number of sessions in $D_k$. \\
$z(D_k)$ & Metadata summary of Starlink profile $D_k$. & $\mu_{\mathrm{tp}}$, $q_{0.05}^{\mathrm{tp}}$, $\sigma_{\mathrm{tp}}$ & Mean throughput, 5th-percentile throughput, and throughput volatility. \\
$\mu_{\mathrm{lat}}$, $H_{\mathrm{ho}}$ & Mean latency and handover or serving-satellite change intensity. & $\mathcal{M}$, $\mathcal{M}_k$, $\mathcal{S}_{\mathrm{mem}}$ & Regression memory, memory for profile $D_k$, and retained representative sessions. \\
$C$, $C_v$, $C_{v+1}$ & ABR controller, current controller version, and committed next version. & $C_v\oplus p$ & Controller obtained by applying patch $p$ to $C_v$. \\
$\operatorname{Replay}(C,\tau)$, $\ell(C,\tau)$ & Replay interface and replay log. & $Q(C,\tau)$, $B(C,\tau)$ & Session QoE and cumulative rebuffering under controller $C$. \\
$a_t$, $u(a_t)$, $\rho_t$ & Bitrate action, bitrate utility, and chunk rebuffering. & $\mu$, $\eta$ & Rebuffering and bitrate-smoothness penalty weights. \\
$\bar Q(C,D)$, $\bar B(C,D)$ & Profile-level mean QoE and mean cumulative rebuffering. & $R_{\mathrm{sev}}(C,D)$ & Severe-session ratio of controller $C$ on profile $D$. \\
$T_{\alpha}(C,D)$, $\operatorname{Tail}_{\alpha}(C,D)$ & Worst-tail rebuffering metric and the worst $\alpha$ fraction of sessions. & $b_{\mathrm{sev}}$, $\alpha$ & Severe-stall threshold and tail fraction. \\
$p$, $\mathcal{P}$, $\Omega(p)$ & Candidate patch, admissible patch space, and patch complexity. & $\operatorname{Schema}_{C_v}(p)$ & Patch schema validation test for controller version $C_v$. \\
$\mathcal{I}_{\mathrm{trig}}$ & Maintenance-trigger indicator. & $\beta_{\mathrm{sev}}$, $\beta_{\mathrm{tail}}$ & Trigger thresholds for severe-session ratio and worst-tail rebuffering. \\
$s(C,D)$, $\mathcal{X}_k$ & Failure summary and maintenance context for profile $D_k$. & $N_{\mathrm{sev}}$, $\rho_{\mathrm{audit}}$ & Number of severe sessions and audit-intervention rate. \\
$\mathcal{F}_{\mathrm{sev}}$, $\mathcal{F}_{\mathrm{tail}}$ & Severe-session set and worst-tail session set. & $G_{\theta}$, $\mathcal{H}_v$ & Agentic patch proposer and Patch History. \\
$\mathrm{meta}(C_v)$ & Controller metadata supplied to the proposer. & $\Gamma$, $\Gamma_{\mathrm{tar}}$, $\Gamma_{\mathrm{mem}}$ & Verifier constraints and their target-profile and regression-memory blocks. \\
$D_{\mathrm{new}}$ & Newly exposed target profile in the verifier constraints. & $\delta$, $\gamma$ & Required severe-risk and tail-risk improvements. \\
$\epsilon$, $\epsilon_{\mathrm{reg}}$ & Allowed QoE loss on the target profile and regression memory. & $\zeta$, $\kappa$ & Allowed severe-risk increase on regression memory and patch-complexity budget. \\
\bottomrule
\end{tabular}
\end{table*}

\section{Agent Prompt and Patch Details}
\label{app:agent-prompt}

This supplementary material records the proposer interface in SafeLink-Agent. It specifies the structured context provided to the proposer, the prompt template, and representative candidate patch outputs. These details support the reproduction of the proposal-generation step used in the maintenance workflow.

\subsection{Structured Prompt Context}

The prompt uses compact summaries instead of full raw traces. When multiple proposal rounds are needed, SafeLink-Agent reuses the same template and updates Patch History with the accepted or rejected patches and their verifier feedback. Table~\ref{tab:appendix-prompt-context} gives an example of the structured context seen by the agentic patch proposer. The concrete values are compacted from replay statistics, controller metadata, patch schemas, verifier constraints, regression memory, and Patch History.

\begin{table*}[!t]
\centering
\caption{Example structured context provided to the agentic patch proposer.}
\label{tab:appendix-prompt-context}
\footnotesize
\begin{tabular}{p{0.20\linewidth}p{0.72\linewidth}}
\toprule
\textbf{Context Field} & \textbf{Example Value Provided to the Agent} \\
\midrule
Profile summary & \texttt{profile\_id}: \texttt{osn\_low\_tail}; \texttt{region}: OSN; \texttt{num\_sessions}: 173; \texttt{p05\_throughput}: 63.8 Mbps; \texttt{latency\_p95}: 78 ms; \texttt{tags}: [low-tail, high-latency]. \\
Failure summary & \texttt{controller}: RobustMPC-v0; \texttt{session\_gt10s}: 2.60\%; \texttt{worst5\_rebuf}: 11.77 s; \texttt{symptoms}: [aggressive high-bitrate request, buffer depletion after throughput drop]. \\
Controller metadata & \texttt{class}: RobustMPC; \texttt{version}: v0; \texttt{allowed\_knobs}: horizon $\in\{4,5,6\}$, margin $\in\{0.80,0.85,0.90\}$, bitrate cap $\in\{30,60,120\}$ Mbps. \\
Patch schema & \texttt{allowed\_type}: profile-conditioned rule; \texttt{operation}: select admissible configuration; \texttt{activation\_features}: [throughput tail, volatility, latency]; \texttt{max\_rules}: 2. \\
Verifier constraints & \texttt{target\_session\_gt10s}: $\leq$ 1\%; \texttt{target\_qoe\_loss}: $\leq$ 8\%; \texttt{memory\_qoe\_loss}: $\leq$ 2\%; \texttt{memory\_session\_increase}: $\leq$ 0.5 pp; \texttt{memory\_worst5\_increase}: $\leq$ 1 s. \\
Regression memory & \texttt{memory\_profiles}: [initial, high-volatility]; \texttt{reference\_qoe}: 96.58; \texttt{reference\_session\_gt10s}: 0.00\%; \texttt{anchor\_sessions}: 96. \\
Patch History & \texttt{accepted}: low-tail rule \#3; \texttt{rejected}: global-safe rule \#2; \texttt{last\_reject\_reason}: QoE loss on regression memory = 8.05\%. \\
\bottomrule
\end{tabular}
\end{table*}

\subsection{Prompt Template}

The prompt is organized as an instruction-following maintenance task. In our implementation, DeepSeek-V4-Pro receives a system instruction, structured context, patch schema, verifier constraints, and output-format requirement. The same template is reused across proposal rounds, while Patch History provides the feedback from previous replay-verification results. The template below shows the structure.

{\footnotesize
\begin{verbatim}
SYSTEM:
You are a controller maintenance proposer for ABR.
Suggest only candidate patches
allowed by the schema.
Do not output per-chunk bitrate actions.

TASK:
Given a new Starlink profile and replay evidence,
propose one patch that reduces severe sessions while
avoiding QoE regression on memory profiles.

INPUT:
1) profile_summary
2) failure_summary
3) controller_metadata
4) patch_schema
5) verifier_constraints
6) regression_memory_summary
7) patch_history

CONSTRAINTS:
- Use only approved patch type and features.
- Keep all parameters inside candidate bounds.
- Satisfy target-profile risk and QoE limits.
- Satisfy regression-memory limits.
- Prefer localized profile-conditioned repairs.
- Explain expected target improvement and risk.

OUTPUT:
Return one candidate patch and rationale.
\end{verbatim}
}

\subsection{Patch Records}
\label{app:patch-records}

The proposer output is parsed into a JavaScript Object Notation (JSON)-like record with typed fields. Table~\ref{tab:appendix-candidate-patch-records} gives compact examples of candidate patch records across the controller classes considered in this paper. These examples make explicit that the proposer output names the controller, states when the patch is activated, selects an allowed operation, and fills only bounded parameters.

\begin{table*}[!t]
\centering
\caption{Example candidate patch records generated by the agentic patch proposer.}
\label{tab:appendix-candidate-patch-records}
\footnotesize
\resizebox{\textwidth}{!}{%
\begin{tabular}{p{0.13\linewidth}p{0.18\linewidth}p{0.25\linewidth}p{0.22\linewidth}p{0.14\linewidth}}
\toprule
\textbf{Controller} & \textbf{Patch Type} & \textbf{Activation Condition} & \textbf{Operation} & \textbf{Bounded Parameters} \\
\midrule
BOLA & Profile-conditioned configuration & Activate on low-tail or high-volatility profiles with severe-session evidence. & Select a buffer-threshold configuration for the matching profile. & Lower buffer $=15$ s; upper buffer $=45$ s. \\
RobustMPC & Profile-conditioned configuration & Activate on low-tail or high-latency profiles with repeated buffer depletion. & Select prediction horizon, safety margin, and bitrate cap from admissible candidates. & Horizon $=5$; margin $=0.85$; bitrate cap $=60$ Mbps. \\
Learned ABR & Runtime auditor patch & Activate when low buffer and low safe capacity make the requested bitrate risky. & Attach or tune a runtime auditor while leaving the learned policy unchanged. & Safe-capacity margin $=0.90$; buffer guard $=1.0$ s. \\
\bottomrule
\end{tabular}
}
\end{table*}

Table~\ref{tab:appendix-patch-output} shows a concrete RobustMPC patch output in the same record style. A candidate that misses a required field, uses an unsupported operation, or violates parameter bounds is rejected before replay.

\begin{table*}[!t]
\centering
\caption{Example RobustMPC patch output from the agentic patch proposer.}
\label{tab:appendix-patch-output}
\footnotesize
\begin{tabular}{p{0.22\linewidth}p{0.70\linewidth}}
\toprule
\textbf{Patch Field} & \textbf{Example Output Value} \\
\midrule
\texttt{type} & \texttt{profile\_conditioned\_rule} \\
\texttt{target\_controller} & \texttt{RobustMPC-v0} \\
\texttt{condition} & Activate when \texttt{profile.tag} contains low-tail or high-latency and \texttt{p05\_throughput} $<$ 65 Mbps. \\
\texttt{operation} & Select an admissible safe configuration for the matching profile; do not change per-chunk bitrate actions. \\
\texttt{parameters} & \texttt{horizon}: 5; \texttt{safety\_margin}: 0.85; \texttt{bitrate\_cap}: 60 Mbps; \texttt{max\_activation\_rules}: 1. \\
\texttt{verifier\_goals} & Reduce \texttt{session\_gt10s} on the target profile; keep QoE loss on regression memory $\leq$ 0.5\%; do not increase memory severe-session ratio. \\
\texttt{rationale} & The target profile shows a weak throughput tail and repeated buffer-depletion events after high-bitrate requests. The patch should reduce aggressive requests only on matching profiles. \\
\bottomrule
\end{tabular}
\end{table*}

\end{document}